\documentclass[useAMS,usegraphicx,usenatbib]{mn2e}
\usepackage{times}   
\usepackage{mathptm} 
\usepackage{color}
\usepackage{xcolor}
\usepackage{amsmath}
\usepackage{url}
\usepackage[normalem]{ulem}
\title[Particle hydrodynamics with accurate gradients]
{Particle hydrodynamics with accurate gradients: a comparison of different formulations}
\author[Rosswog]
{S. Rosswog\thanks{E-mail: stephan.rosswog@astro.su.se}$^{1,2}$\\
$^1$ University of Hamburg, Hamburger Sternwarte, Gojenbergsweg 112, 21029, Hamburg, Germany\\
$^2$ The Oskar Klein Centre, Department of Astronomy, AlbaNova, Stockholm University, SE-106 91 Stockholm, Sweden}

\def\p{\partial}

\def\be{\begin{equation}}
\def\ee{\end{equation}}
\def\bi{\begin{itemize}}
\def\i{\item}
\def\ei{\end{itemize}}
\def\ben{\begin{enumerate}}
\def\een{\end{enumerate}}
\def\bea{\begin{eqnarray}}
\def\eea{\end{eqnarray}}
\def\bt{\begin{tabbing}}
\def\et{\end{tabbing}}

\def\edo{\end{document}}

\def\f{{\bf f}}

\def\M4{M$_4$}
\def\M6{M$_6$}
\def\lcm6{LCM$_6$}
\def\qcm6{QCM$_6$}

\def\wh3{W$_{\rm H,3}$}
\def\wh4{W$_{\rm H,4}$}
\def\wh5{W$_{\rm H,5}$}
\def\wh6{W$_{\rm H,6}$}
\def\wh7{W$_{\rm H,7}$}

\newcommand{\Ma}{\texttt{MAGMA2}\;}

\def\vr{\vec{r}}

\begin{document}
\date{Draft version}

\pagerange{\pageref{firstpage}--\pageref{lastpage}} \pubyear{2024}

\maketitle

\label{firstpage}

\begin{abstract}
We compare here several modern versions of SPH with a particular focus on the impact
of gradient accuracy. We examine specifically an approximation to the "linearly exact"
gradients (aLE) with standard SPH kernel gradients and with linearly reproducing kernels (RPKs)
that fulfill the lowest order consistency relations exactly by construction. Most of the explored 
SPH formulations use shock dissipation (i.e. artificial viscosity and conductivity) with slope-limited 
reconstruction and parameters that trigger on both shocks and noise. We also compare 
with a recent particle hydrodynamics formulation that uses both RPKs and Roe's approximate Riemann solver 
instead of shock dissipation.  Not too surprisingly, we find that the shock tests are rather insensitive to the gradient accuracy, 
but whenever instabilities are involved the gradient accuracy plays a crucial role. The reproducing 
kernel gradients perform best, but they are closely followed by the much simpler aLE gradients. 
The Riemann solver approach has some (minor) advantages in the shock tests, but shows some resistance 
against instability growth and at low resolution the corresponding Kelvin-Helmholtz simulations show substantially slower 
instability growth than the best shock dissipation approaches. Based on the battery of benchmark tests
performed here, we consider a shock dissipation approach with reproducing kernels (our versions $V_3$
and $V_5$) as best, but closely followed by a similar version ($V_2$) that uses the simpler and computationally 
cheaper aLE gradients.
\end{abstract}

\begin{keywords}
hydrodynamics -- methods: numerical -- instabilities -- shock waves -- software: simulations
\end{keywords}

\section{Introduction}
\label{sec:intro}
Smooth particle hydrodynamics (SPH) is the "archetypical" mesh-free hydrodynamics method
of astrophysical gas dynamics \citep{lucy77,gingold77,benz90a,monaghan92}. 
It has a number of benefits that include  its natural adaptivity, the ease with which it treats
vacuum (without needing an artificial background medium or "atmosphere"), its excellent advection 
properties and its Galilean invariance. For a broad overview over SPH we refer to a number
extensive review articles \citep{monaghan05,rosswog09b,price12a,rosswog15c,rosswog26a}. 
Simulating particle-based gas flows  is still a relatively young field compared to Eulerian gas dynamics, 
and much development is still ongoing, both in terms of improving the methodology and in terms of 
spreading into new research areas. While having originated from (non-relativistic) astrophysics,
SPH has by now found many industrial applications, see e.g. \cite{monaghan12a} and \cite{quinlan21}.
On the astrophysics side, the last years have seen much development towards relativistic particle
hydrodynamics, either in special relativity \citep{rosswog10b,rosswog11a,rosswog15b,kitajima26},
in general relativistic hydrodynamics in a fixed metric \citep{rosswog10a,tejeda17a,liptai19}, or, more recently, in fully dynamical
numerical relativity, see \cite{rosswog21a,diener22a,rosswog23a,magnall23} and the dedicated book chapter  in
\cite{rosswog25c}.\\
One of SPH's pivotal  strengths is that it can be formulated in way so that mass, energy, momentum 
and angular momentum are conserved exactly\footnote{This is true if the time integration is exact, in practice time integration
and fast, but approximate treatment of forces can lead to small non-conservation effects.}. The straight
forward exact conservation is in standard SPH guaranteed by the anti-symmetries of the kernel gradients
$\nabla_a W_{ab}= - \nabla_b W_{ab}$ (ensuring exact conservation of energy and momentum) and
by kernel gradients pointing exactly along the line connecting two particles $\nabla_a W_{ab} \propto \hat{e}_{ab}$
(ensuring angular momentum conservation).\\
It has been known for some time that gradient estimates that are based  on direct gradients of the smoothing 
kernel are not particularly accurate, see e.g. \cite{abel11,garcia_senz12,rosswog15b,frontiere17}, and one may 
wonder whether sacrificing exact angular momentum conservation in favor of more 
accurate gradients may overall be a better compromise. Many accurate mesh-less gradient prescriptions exist, see Sec. 2.7
in \cite{rosswog26a}, but to our knowledge none of them can guarantee exact angular momentum
conservation {\em and} exactly reproduce the gradients of linear functions. There are several approaches
where {\em exact} angular momentum conservation was sacrificed in favor of good gradient accuracy
\citep{garcia_senz12,frontiere17,rosswog25a,sandnes25} and all these studies found favorable results
for the more accurate gradients.
Where it has been quantified, see for example \cite{frontiere17} or Fig.~30 in \cite{rosswog20a}, the non-conservation 
has been found to be well below 1\%,  and  -at least in the study of \cite{rosswog20a} -- this small non-conservation
is  dominated by approximate gravitational forces via a tree-approach  rather than the gradient
prescription.\\
In this paper we are particularly interested in 
\ben
\i[a)] the effects of improved gradient accuracy in standard
hydrodynamic benchmark tests, 
\i[b)] a comparison between simple gradient correction methods that only 
need the inversion of a 3$\times3$ matrix (Sec.~\ref{sec:aLE}) with the reproducing kernel (RPK) method 
that ensures an exact partition of unity/nullity  (Sec.~\ref{sec:RPK}), but is also computationally more
expensive and
\i[c)] how well a shock dissipation approach with RPKs 
performs in comparison to a recently suggested RPK method \citep{rosswog25a} that uses 
Roe's approximate Riemann solver \citep{roe86}.
\een
This paper is structured as follows. In Sec.~\ref{sec:SPH_approx} we concisely summarize the basic SPH approximation,
the conditions that need to be fulfilled for high accuracy and we discuss several approaches to approximate gradients at the particle
positions. Apart from standard SPH kernel gradients, we discuss an approximation to "linearly exact gradients"
that has to our knowledge not been used in astrophysics  and we summarize the construction of RPKs and of their gradients. 
In Sec.~\ref{sec:shock_diss} we summarize our "shock dissipation", i.e. artificial viscosity and conductivity, terms. 
Here we enhance a \cite{cullen10} type shock trigger with a sensitive trigger on noise that is based
\cite{rosswog15b}. We then formulate in Sec.~\ref{sec:SPH_formulations} five different SPH versions
the abilities of which are explored in the tests described in Sec.~\ref{sec:results}. In this section we explore shocks, instabilities 
and complex Riemann problems due to Schulz-Rinne. We finally summarize our findings 
in Sec.~\ref{sec:summary}.

\section{Methodology}
\label{sec:method}

\subsection{SPH approximation}
\label{sec:SPH_approx}
The standard SPH way of smoothly approximating a  function $f$ that is known at particle positions $b$
is
\be
\tilde{f}(\vr)= \sum_b V_b f_b W_h(\vr - \vr_b) \equiv \sum_b f_b \Phi^h_b(\vr),
\label{eq:SPH_approx}
\ee 
where for the particle volume $V_b$ one usually uses $\frac{m_b}{\rho_b}$, $W_h$ is the SPH smoothing kernel
whose support size is determined by the smoothing length $h$, which can be considered as a function
of position, i.e. $h = h(\vr)$.  We have further 
introduced the shape function $\Phi^h_b(\vr)= V_b W_h(\vr - \vr_b)$.  To quantitatively understand the accuracy 
of this approximation, insert a Taylor expansion of $f_b$ around $\vr_a$
\be
f_b= f_a + (\p_i f)_a (\vr_{ba})^i + \frac{1}{2!} (\p_{jk} f)_a (\vr_{ba})^j (\vr_{ba})^k + ... 
\label{eq:Taylor_fb_at_a}
\ee
with $\vr_{ba}= \vr_b - \vr_a$ into Eq.~(\ref{eq:SPH_approx}) and compare the resulting expression 
with Eq.~(\ref{eq:SPH_approx}), evaluated at $\vr = \vr_a$. Requiring that $\tilde{f}(\vr_a)$ is a good approximation 
to $f_a$ immediately provides the following conditions on the discrete moments
\bea
(M_0)_a       &=& \sum_b \Phi_b^h(\vr_a) \stackrel{!}{=} 1 \label{eq:zeroth_moment}\\
(M_1^i)_a    &=& \sum_b \vr_{ba}^i \; \Phi_b^h(\vr_a) \stackrel{!}{=} 0 \label{eq:first_moment}\\
(M_2^{jk})_a &=& \sum_b \vr_{ba}^j \vr_{ba}^k  \; \Phi_b^h(\vr_a) \stackrel{!}{=} 0 \label{eq:second_moment}\\
...\nonumber
\eea
for an accurate approximation. So to reproduce a constant function exactly, the condition (\ref{eq:zeroth_moment})
needs to be fulfilled, to reproduce a linear function exactly  condition (\ref{eq:first_moment}) needs to be fulfilled and
so on. The condition on $M_0$ is of course just the requirement of an exact  "partition of unity" 
and for later use we write it explicitly, together with its gradient, as
\be
\sum_b V_b \; W_h(r_{ab}) = 1 \quad {\rm and} \quad \sum_b V_b \; \nabla W_h(r_{ab}) = 0.
\label{eq:PoU}
\ee
The latter condition is sometimes referred to as "partition of nullity".
In standard SPH these "zero order consistency relations" are not enforced and only approximately fulfilled.

\subsection{Gradients in SPH}
The conventional way to calculate SPH gradients is to 
straight forwardly apply the nabla operator to expression Eq.~(\ref{eq:SPH_approx})
\be
\nabla f (\vr)= \sum_b V_b f_b \nabla W_h(\vr - \vr_b),
\ee
or, when specifying to the gradient at the position of a particle $a$, $\vr_a$,
\be
(\nabla f)_a= \sum_b V_b f_b \nabla W_{h_a}(\vr_a - \vr_b).
\label{eq:nabla_f_a}
\ee 
For the radial kernels that are usually used in SPH, $W_h(\vr_a - \vr_b)= W_h(|\vr_a - \vr_b|)$,
this has the virtue of anti-symmetry, 
\be
\nabla_a W_h(|\vr_a - \vr_b|)= - \nabla_b W_h(|\vr_a - \vr_b|),
\ee
which allows in a straight-forward way to ensure the conservation of momentum and energy in SPH.
Moreover, since the kernel gradient points along the line connecting particles $a$ and $b$,  
\be
\nabla_a W_h(|\vr - \vr_b|) \propto \vr_a - \vr_b
\label{eq:nablaW_prop_rab}
\ee
angular momentum can be easily conserved by
construction, see e.g. Sec. 2.3 in \cite{rosswog26a} for a recent review.\\
This straight forward enforcement of exact conservation is a very good reason to use the above gradient
prescription, but it comes at the price of only approximately fulfilling the lowest order interpolation 
consistency relations and such gradients are not very accurate, see below.
How well the consistency conditions are fulfilled in practice depends usually on the local particle distribution,
which in turn depends on the combination of kernel function and contributing neighbors. While
the consistency relations can be fulfilled to good accuracy e.g. by the combination of Wendland kernels \citep{wendland95}
with many neighbor particles (i.e. a large smoothing length), this comes at the price
of long neighbor lists which require a non-negligible computational burden. As we will show below,
see Fig.~\ref{fig:gradient_experiment}, even a seemingly small deviation
from a regular particle distribution can lead to a serious loss of gradient accuracy.\\
There are, however, many ways to calculate accurate gradients at particle positions, see Sec. 2.7 in \cite{rosswog26a} 
for an overview, though the gained gradient accuracy usually comes at the price of sacrificing property Eq.~(\ref{eq:nablaW_prop_rab}) and thus losing 
exact angular momentum conservation. Here we want to explore in particular  an approximation to the linearly exact
(LE) gradients, see Sec.~\ref{sec:aLE} and "reproducing kernel gradients" \citep{liu95}, which are derived by explicitly enforcing the consistency 
relations up  to linear order, see Sec.~\ref{sec:RPK}.

\subsubsection{Linearly exact gradients and their approximations}
\label{sec:aLE}
Linearly exact (LE) gradients can be obtained \citep{price04c} by inserting 
Eq.~(\ref{eq:Taylor_fb_at_a}) into the RHS of Eq.~(\ref{eq:nabla_f_a})
\be
\sum_b V_b  f_b \nabla_a W_h(r_{ab})= \sum_b V_b \left\{ f_a + (\nabla f)_a \cdot (\vr_b - \vr_a) + ...  \right\} \nabla_a W_h(r_{ab}).
\ee
Rearranging and solving  for the gradient yields
\be
\left(\nabla^i f\right)^{\rm LE}_a= C_{a}^{ik} \sum_b V_b (f_b-f_a)
\nabla_a^k W_{h_a}(r_{ab})= \sum_b V_b f_{ba} \widetilde{\nabla^i_a W}_{h_a}(r_{ab}),
\label{eq:lin_exact_gradient}
\ee
where $f_{ba}= f_b - f_a$, the ``corrected kernel gradient'' reads
\be
\widetilde{\nabla^i_a W}_{h_a}(r_{ab})=  C_{a}^{ik}  \; \nabla_a^k W_{h_a}(r_{ab})
\ee
and the correction matrix is given by
\be
C_a^{ik}= \left( \sum_b  V_b \left(\vr_b - \vr_a \right)^i \nabla_a^k W_{h_a}(r_{ab})\right)^{-1}.
\label{eq:correction_matrix}
\ee
If we assume that we have to a good accuracy a partition of unity, see the second relation in 
Eq.~(\ref{eq:PoU}), we can neglect the term containing $f_a$ in Eq.~(\ref{eq:lin_exact_gradient}) 
and approximate the gradient as 
\be
\left(\nabla^i f\right)^{\rm aLE}_a = C_a^{ik} \sum_b V_b f_b \nabla^k_a W_{h_a}(r_{ab}). 
\label{eq:aLE_gradient}
\ee
The comparison of Eq.~(\ref{eq:aLE_gradient}) with Eq.~(\ref{eq:nabla_f_a}) proposes
\be
\nabla_a^i W_{h_a}(r_{ab}) \rightarrow C_a^{ik} \nabla_a^k W_{h_a}(r_{ab}) \; {\rm and}  \;
\nabla_b^i W_{h_b}(r_{ba}) \rightarrow C_b^{ik} \nabla_b^k W_{h_b}(r_{ba}),
\label{eq:aLE_replacement}
\ee
where the second relation, which is needed below, results simply from exchanging $a \leftrightarrow b$.\\
As pointed out in \cite{rosswog15b}, Eq.~(\ref{eq:lin_exact_gradient}) is equivalent to an approach   
that was derived by discretizing an integral 
\citep{garcia_senz12,cabezon12a}. Their resulting "integral approximation" (IA) gradient prescription reads
\bea
(\nabla^i f)^{IA}_a&=& D_{a}^{ik} \; \sum_b V_b \; (f_b-f_a) \; \vr_{ba}^k  W_{h_a} (\vr_{ab}),\\
&\approx& D_{a}^{ik} \; \sum_b V_b \; f_b \vr_{ba}^k W_{h_a} (\vr_{ab}),
\label{eq:IA_gradient}
\eea
where their correction matrix is 
\be
D_{a}^{ik}= \left( \sum_b V_b (\vr_a - \vr_b)^i (\vr_a - \vr_b)^k W_{h_a}(r_{ab} )\right)^{-1}
\label{eq:IA_correction_matrix}
\ee
and when neglecting the $f_a$-term in Eq.~(\ref{eq:IA_gradient})
we have assumed that Eq.~(\ref{eq:first_moment}) is fulfilled. 
To see that this approach is equivalent to Eq.~(\ref{eq:lin_exact_gradient}), 
start from
\be
\nabla_a W_{h_a}(\vec{r}_{ab})= \frac{\p}{\p \vr_a}W\left(\frac{|\vr_a
    - \vr_b|}{h_a}\right)= \frac{\p W}{\p u} \frac{\p u}{\p \vr_a}=
\frac{\p W}{\p u}  \frac{\vr_a -\vr_b}{h_a|\vr_a - \vr_b|},
\ee
where $u\equiv |\vr_a - \vr_b|/h_a$.  Since $\p W/\p u <0$, we can
write this as
\be
\nabla_a W_{h_a}(\vec{r}_{ab})= - \frac{\p W}{\p u} \frac{\vr_b
  -\vr_a}{h_a|\vr_a - \vr_b|} \equiv (\vr_b -\vr_a)
\tilde{W}_{h_a}(\vr_{ab}),
\label{eq:gradW_tildeW}
\ee
where $\tilde{W}$ is another valid, positive definitive kernel with finite
support. Inserting Eq.~(\ref{eq:gradW_tildeW}) into the LE-gradient Eq.~(\ref{eq:lin_exact_gradient})
we find
\begin{align}
(\nabla_a^i f)^{IA}= & \left( \sum_l V_l (\vr_b
  -\vr_a)^i (\vr_b - \vr_a)^k \tilde{W}_{h_a}(
  \vr_{ab})\right)^{-1} \nonumber\\
  & \left\{ \sum_b V_b (f_b - f_a) (\vr_b -
  \vr_a)^k \tilde{W}_{h_a}(\vr_{ab})\right\},
\label{eq:der_fIA}
\end{align}
which is, apart from using $\tilde{W}$ instead of $W$, the gradient expression found
by discretizing an integral expression.\\
Calculating such correction matrix gradients comes at a very small computational price, since only
a $3 \times 3$ matrix, either Eq.~(\ref{eq:correction_matrix}) or Eq.~(\ref{eq:IA_correction_matrix}), 
needs to be inverted and this can be done analytically.

\subsubsection{Reproducing kernel gradients}
\label{sec:RPK}
The standard SPH-approximation neither exactly reproduces
constant nor linear functions. This, however, can be enforced \citep{liu95}, by enhancing the 
kernel functions with additional parameters $A$ and $B^i$,
\be
\mathcal{W}_{ab}(\vec{r}_{ab}) \equiv A_a \left[1 + B_a^i \; (\vec{r}_{ab})^i\right] \bar{W}_{ab},
\ee
where $\bar{W}_{ab}$ is a symmetrized kernel function, e.g.
\be
\bar{W}_{ab}= \frac{1}{2}  \left[W_{h_a}(r_{ab})+ W_{h_b}(r_{ab})\right]. \label{eq:Wbar}
\ee
The parameters $A$ and $B^i$ are determined at every point (here labelled
$a$) so that $\mathcal{W}$ exactly reproduces the discrete first-order consistency relations
\be
\sum_b V_b \; \mathcal{W}_{ab} = 1 \quad {\rm and} \quad
\sum_b (\vec{r}_{ab})^i \;  V_b \; \mathcal{W}_{ab} = 0,
\label{eq:consistency_relations}
\ee
see the moment conditions Eqs.~(\ref{eq:zeroth_moment}) and (\ref{eq:first_moment}).  The
price for this exact reproduction is that the kernel is, apart from the
computational effort to determine $A$ and $B^i$,  no longer
guaranteed to be radial due to the $B^i$-term. This property is in
standard SPH responsible for exact angular momentum conservation.  It
is, however, possible to relatively straight-forwardly write a set of
equations that conserves energy, momentum and mass, if density
summation is used, but not necessarily angular momentum, since
the mutual accelerations  can no longer be guaranteed to be along
the line connecting two particles, see
Sect.~2.3 in \cite{rosswog26a}. \\
\begin{figure*}
  \centerline{
    \includegraphics[width=7cm]{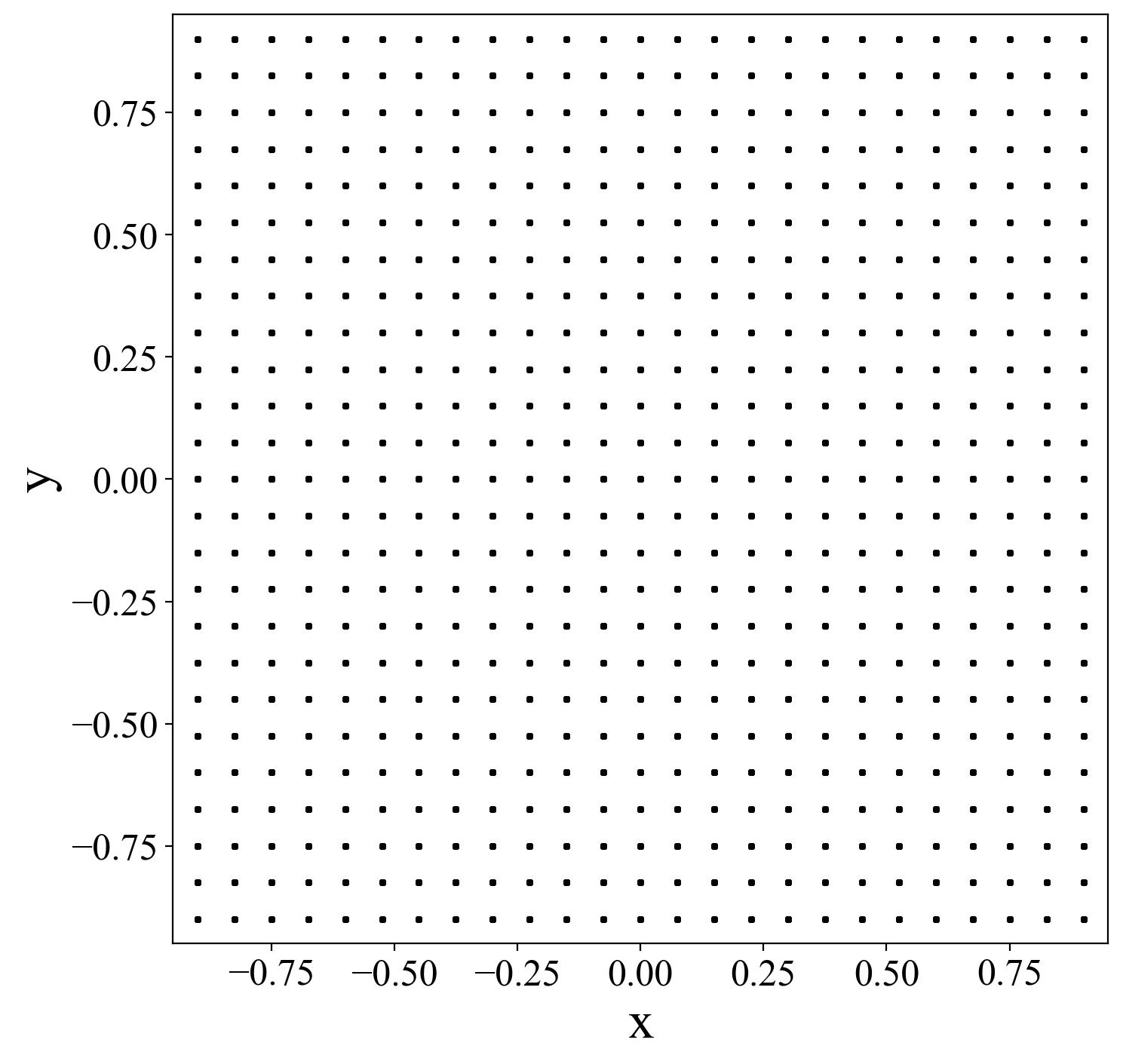}
    \includegraphics[width=7cm]{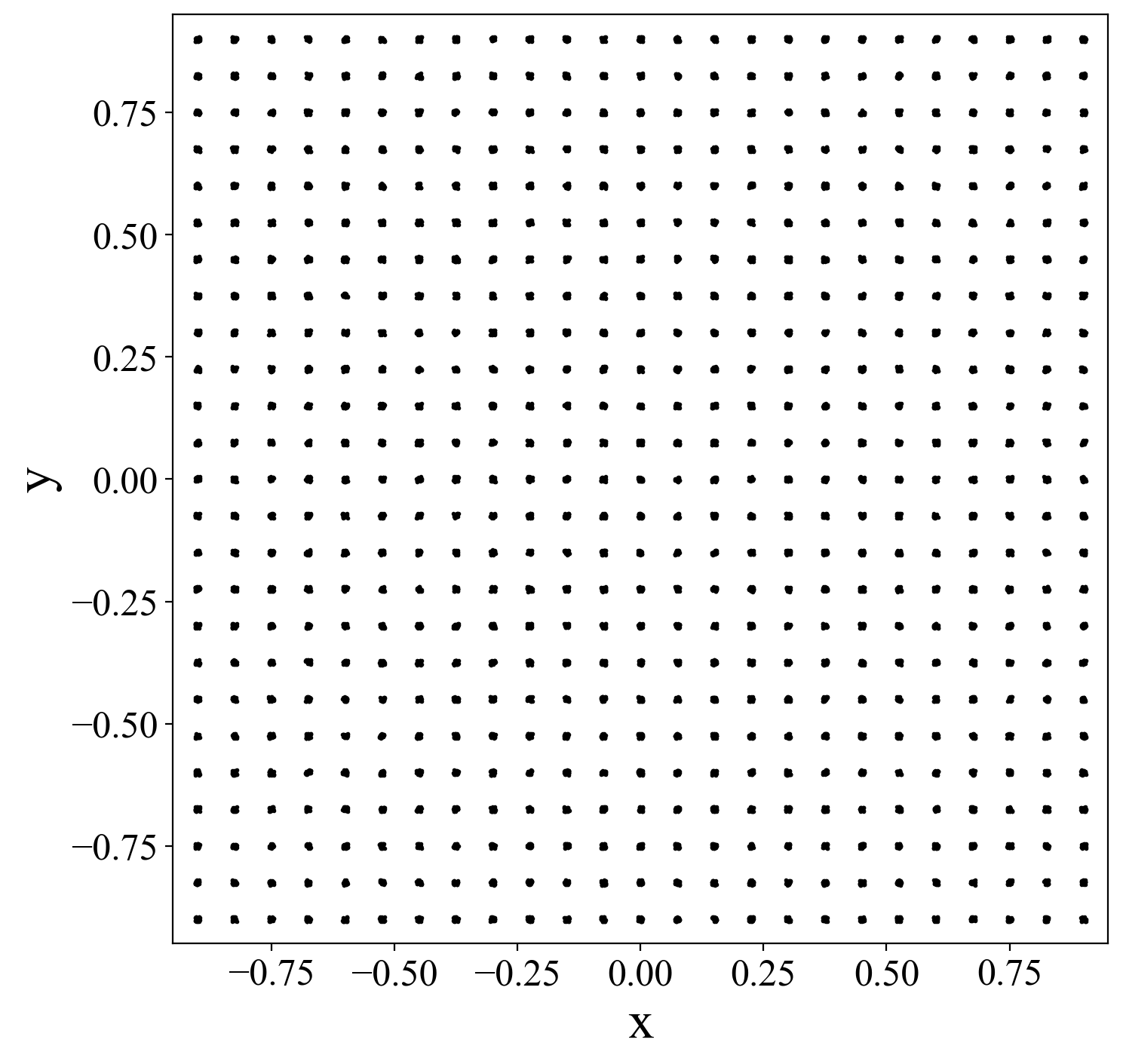}
  }
  \centerline{
    \includegraphics[width=7cm]{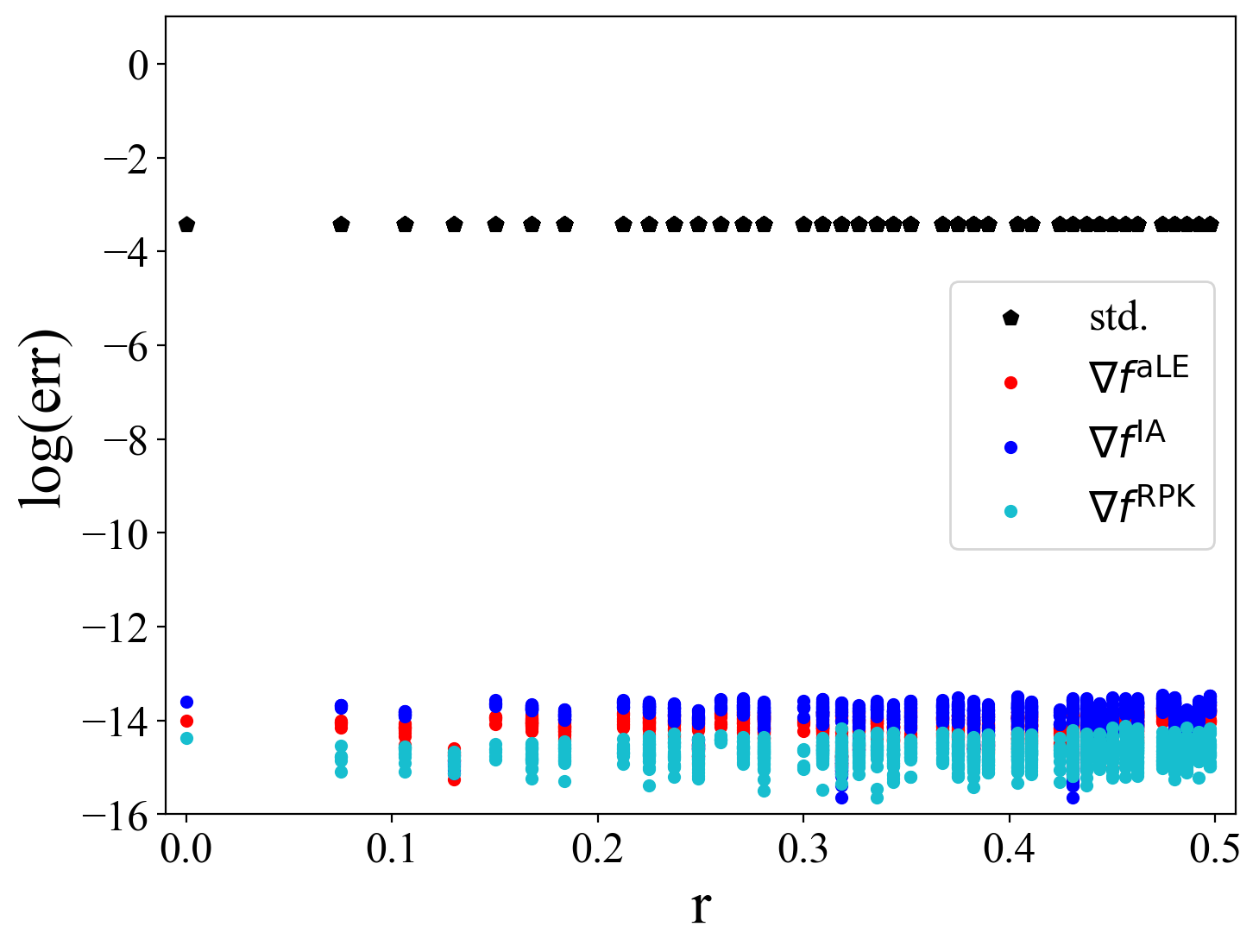}
    \includegraphics[width=7cm]{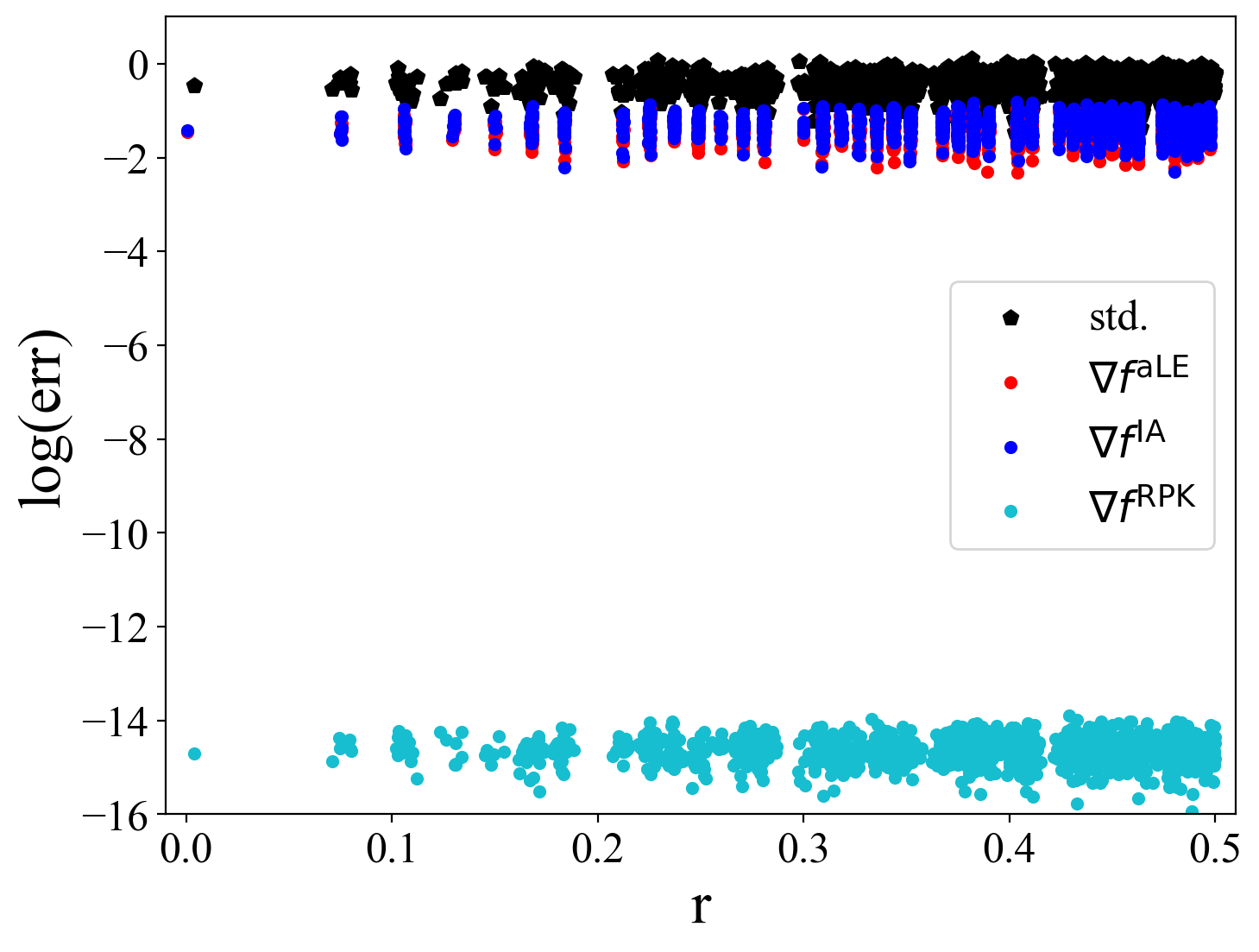}
  }
 \caption{Numerical experiment to measure gradient accuracy. The upper
 row shows the particle positions, exactly arranged on a cubic
 lattice (left), and the same lattice, but with each particle coordinate
shifted by a small random number chosen uniformly from the interval
[-0.001,0.001].
The second row shows the gradient errors for both particle
configurations for the different gradient gradient prescription. Note
that the hardly visible randomization of the positions leads to a serious
gradient deterioration for the standard, the aLE- and the IA-gradient. Please see the main text for the exact
definitions of the gradient estimates.}
   \label{fig:gradient_experiment}
\end{figure*}
Once the kernels are explicitly constructed, see Appendix \ref{sec:App_RPK}
for the details, one can  approximate a function $f$ and its derivative via
\be
f(\vr_a)= \sum_b V_b \; f_b \; \mathcal{W}_{ab}  \quad
{\rm and} \quad
\nabla^k f_a= \sum_b V_b \; f_b \; \p_k \mathcal{W}_{ab}.
\ee
These expressions look very similar to standard SPH approximations, but they
{\em exactly} reproduce linear functions on a discrete level, which the SPH equations do not.
The gradient of $\bar{W}_{ab}$ can be written as
\be
\nabla_a \bar{W}_{ab}= \frac{ \nabla_a W_{h_a}(\vr_{ab}) + \nabla_a
  W_{h_b}(\vr_{ab}) }{2}  
  = \frac{ \nabla_a W_{h_a}(\vr_{ab}) - \nabla_b
  W_{h_b}(\vr_{ab})}{2},
  \label{eq:nabla_barW}
\ee
which suggests the replacement
\be
\nabla_a \bar{W}_{ab} \rightarrow (\nabla \mathcal{W})_{ab} \equiv
\frac{1}{2} \left[ \nabla_a \mathcal{W}_{ab} - \nabla_b
  \mathcal{W}_{ba}\right].
\label{eq:RPK_gradient}
\ee
Various particle hydrodynamics schemes based on reproducing
kernels have been developed in recent times, for example the artificial
viscosity-based  codes \texttt{CRKSPH}
\citep{frontiere17} and \texttt{REMIX} \citep{sandnes25} or a
Riemann solver-based approach with reproducing kernels which is described 
in \citet{rosswog25a}.\\

\subsubsection{Measuring gradient accuracy}
We show in Fig.~\ref{fig:gradient_experiment}  an experiment to scrutinize the gradient
approximations. In a first step, we use  particles that are exactly arranged on
a cubic lattice and, in a second step,  this lattice is
randomly disturbed by a small amplitude, see the upper two panels of
Fig.~\ref{fig:gradient_experiment}. In the randomized case, right column 
Fig.~\ref{fig:gradient_experiment}, each particle coordinate was displaced 
by a random number chosen uniformly from the interval [-0.001,0.001]. 
Each particle is assigned a function $\f(\vr)= (-x,0,0)$ and we measure 
the error between the numerically determined gradient and the exact result, 
$\epsilon\equiv |\nabla f^\mathrm{num} - \nabla f^\mathrm{ex}|/|\nabla f^\mathrm{ex}|$.\\
Note that the hardly visible perturbation (right panel in Fig.~\ref{fig:gradient_experiment}) 
has a serious impact on the accuracy of the standard and the IA- and aLE-gradients, 
see Eqs.~(\ref{eq:nabla_f_a}), (\ref{eq:aLE_gradient}) and (\ref{eq:IA_gradient}),
while the RPK gradient estimate reproduces the theoretical result to
within machine precision whether the particle distribution is randomized or not. The IA- and the aLE-gradients
perform very similar in this experiment, with a small advantage for aLE-gradients. This small
difference, however, may not be relevant in practical dynamical benchmarks. 
Also note that both correction matrix approaches assumed in their derivation 
that a partition of unity is fulfilled to a good approximation. Since we see here
that even small deviations from a regular particle distribution lead to a strong loss of accuracy
and it is difficult to predict how (ir)regular the particle distribution will become in a dynamical simulation,
one may wonder whether such matrix correction approaches are actually useful in practice.
As will be shown below, however, they turn out to deliver clearly better results
than the standard SPH gradients.

\subsection{Shock dissipation}
\label{sec:shock_diss}
Before we come to the explicit SPH formulations we want to summarize
how we treat shocks. 
The most common approach to handle shocks in SPH is via "artificial 
dissipation" in the form of "artificial viscosity" (acting on the momentum
equation), often enhanced by "artificial conductivity" (acting on the energy equation).
We will collectively refer to them as "shock dissipation". The main purposes 
of shock dissipation are i)  to prevent a steepening sound wave from becoming
too steep to be treated numerically, ii) to produce the correct entropy in a
shock and iii) to ensure the Rankine-Hugoniot shock jump conditions, see
\cite{mattsson15} for a wider historical perspective on shock dissipation. Note 
that the comparison with Riemann solvers also suggests to use dissipative terms 
for the continuity equation so that particles can exchange mass in a conservative 
way. This has so far been applied only in a few cases, but there the 
authors find good results \citep{read12,sandnes25}.\\
The major change when implementing artificial viscosity is to enhance the 
physical pressure $P$ by an artificial pressure $Q$, 
\be
P \rightarrow \tilde{P}= P + Q, \label{eq:P+Q}
\ee 
wherever it occurs. We use \citep{monaghan83}
\be
Q_a= \rho_a \left(- \alpha c_{{\rm s},a} \mu_a + \beta \mu_a^2\right),
\label{eq:Qvis}
\ee
where the velocity jump estimate is
\be
\mu_a= {\rm min} \left(0,\frac{(\vec{v}_{ab} \cdot \vec{r}_{ab})h_a}{r_{ab}^2 +
\epsilon^2 h_a^2} \right).
\label{eq:mu_vis}
\ee
Here, $\vec{v}_{ab}= \vec{v}_a - \vec{v}_b$, $\vec{r}_{ab}= \vec{r}_a
- \vec{r}_b$, $r_{ab}^2= \vec{r}_{ab} \cdot \vec{r}_{ab}$ and $\alpha,
\beta$ and $\epsilon$ are numerical parameters with typical
values  1, 2 and $0.1$ and $c_{{\rm s}}$ is the sound speed. The
$\epsilon$-term in the denominator avoids divergence and the
min-function switches off the artificial pressure for receding
particles ($\vec{v}_{ab} \cdot \vec{r}_{ab}>0$). Applying dissipation
only in converging flows is a noteworthy difference between shock
dissipation and standard Riemann solver approaches where the 
dissipation acts in both compressing and expanding flows.\\
We also add a small amount of artificial conductivity \citep{price08a} to energy equation
\be
\left(\frac{du}{dt}\right)_{a, \rm AC}= \alpha_u \sum_b \frac{m_b}{\bar{\rho}_{ab}} 
\sqrt{\frac{|P_a - P_b|}{\bar{\rho}_{ab}}}(u_a - u_b) \hat{e}_{ab}\cdot \nabla W_h(r_{ab}),
\label{eq:conduct}
\ee
with $\alpha_u= 0.05$ which has been found advantageous in shocks and
instabilities \citep{rosswog07c,price08a,valdarnini12}.

\subsubsection{Controlling dissipation}
The straight forward implementation of shock dissipation may lead to unwanted dissipation
where it is not needed, we therefore perform i) a slope-limited linear reconstruction in the 
dissipative terms as in the \Ma code \citep{rosswog20a} and ii) we steer the dissipation parameters $\alpha$ and $\beta= 2 \alpha.$\\
For the slope limited reconstruction, we replace the velocity values at the particle positions
in the $\vec{v}_{ab}$ term in Eq.~(\ref{eq:mu_vis}), by values reconstructed to the mid-point
between the particles both from $a$- and the $b$-side
\be
v_a^{i, \rm rec} = v^i_a - \frac{1}{2} \Psi(\nabla v^i_a,\nabla v^i_b)
\cdot \vec{r}_{ab}  \quad 
v_b^{i, \rm rec} =  v^i_b + \frac{1}{2} \Psi(\nabla v^i_a,\nabla v^i_b)
\cdot \vec{r}_{ab} \label{eq:rec1}.
\ee
Similarly, we replace $u_a - u_b$ in Eq.~(\ref{eq:conduct}) by $u$-values that 
are reconstructed from both sides similar to Eq.~(\ref{eq:rec1}). The quantity $\Psi$ is a slope limiter
for which we use  the \texttt{vanAlbada limiter} \citep{vanAlbada82}
\be
\Psi_\mathrm{vA}(x,y)= \begin{cases} \frac{(x^2 + \epsilon^2) y + (y^2 + \epsilon^2) x}{x^2 + y^2 + 2 \epsilon^2} \quad & {\rm if } \; x y > 0,\\
                                                 0 & \text{otherwise}.
                                               \end{cases}
                                               \label{eq:vanAlbada}
\ee
The limiter is insensitive to the exact value of $\epsilon$, here, we use $\epsilon^2= 10^{-6}$.
In \cite{rosswog25a} we had compared various slope limiters and found particularly good 
results for the \texttt{vanAlbada limiter}. Such slope-limited reconstruction approaches in
shock dissipation terms have turned out to be very efficient in suppressing unwanted dissipation
\citep{christensen90,frontiere17,rosswog20a,rosswog21a,sandnes25}.\\
To reduce dissipation even further, the dissipation parameter $\alpha$ can in addition be made time-dependent 
as originally suggested by \cite{morris97}, so that it rises quickly where needed and decays exponentially otherwise.  
If no higher dissipation parameter value is triggered, see below, $\alpha$ evolves according to
\be
\frac{d\alpha_a}{dt}= - \frac{\alpha_a(t) - \alpha_{\min}}{\tau_a},
\label{eq:alpha_steering_MM}
\ee
where $\alpha_{\min}$ represents a minimum, ``floor'' value 
for the viscosity parameter and $\tau_a$ is an
individual decay time scale at particle $a$. Here we use $\alpha_{\min}= 0$ and $\tau_a= 30 h_a/c_{s,a}$.
If instead a higher dissipation parameter $\alpha_{a,\rm des}$
is needed, $\alpha_a(t)$ is instantaneously replaced with $\alpha_{a,\rm des}$. 
\cite{cullen10} suggested as shock-triggered dissipation parameter
\be
\alpha_{a, \rm shock}= \alpha_{\max} \frac{T^s_a}{T^s_a + c^2_a/h_a^2}, \label{eq:alpha_shock}
\ee
where the shock trigger\footnote{They multiplied $T^s_a$ with an additional limiter which we
omit here for simplicity.},
\be
T^s_a=  \max\left[-\frac{d (\nabla\cdot\vec{v})_a}{dt},0\right] \label{eq:trigger_CD},
\ee
measures a temporarily increasing compression. We found that this trigger works well in shocks, but it
triggers too little dissipation in the challenging Schulz-Rinne tests, see Fig.~\ref{fig:effect_noise_trigger}
in the Appendix. We therefore add an additional
trigger on noise that measures the strength of local sign fluctuations in $\nabla \cdot \vec{v}$ \citep{rosswog15b}. To this end we calculate
average $\nabla \cdot \vec{v}$ values separately for positive and negative signs:
\be
\mathcal{S}^+_a= \frac{1}{N^+} \sum_{b,\nabla\cdot \vec{v}_b>0}^{N^+} \nabla \cdot \vec{v}_b
\quad {\rm and} \quad
\mathcal{S}^-_a= - \frac{1}{N^-} \sum_{b,\nabla\cdot \vec{v}_b<0}^{N^-} \nabla \cdot \vec{v}_b
\ee
where $N^+/N^-$ are the correspondingly contributing particle numbers in the neighborhood of particle $a$. 
The noise trigger then reads
\be
T^n_a= \sqrt{ \mathcal{S}_a^+  \mathcal{S}_a^-}
\label{eq:noise_trigger}
\ee
and it is translated to an  $\alpha$-value via
\be
\alpha_{a, \rm noise}= \alpha_{\max} \frac{T^n_a}{T^n_a + 0.01 c_a/h_a}. \label{eq:alpha_noise}
\ee
The final desired dissipation parameter value is then
\be
\alpha_a^{\rm des}= \rm{max}(\alpha_{a, \rm shock},\alpha_{a, \rm noise})
\ee
and if it exceeds $\alpha_a(t)$ the latter is immediately replaced by $\alpha_a^{\rm des}$.
If there are sign fluctuations, but they are small compared to the reference value in the denominator of
Eq.~(\ref{eq:alpha_noise}), the product is
very small.  If instead we have uniform expansion or compression in the neighborhood of particle $a$,
either $\mathcal{S}_a^+$ or  $\mathcal{S}_a^-$ will be zero. So only for sign changes and significantly
large compressions/expansions will the $T^n$ have a substantial
value. Note that we have chosen a small reference value ($0.01 c_a/h_a$) in Eq.~(\ref{eq:alpha_noise}),
which we found to work well in detecting noise in Schulz-Rinne tests without triggering too much dissipation
in rather smooth flows such as Rayleigh-Taylor or Kelvin-Helmholtz instabilities, see 
Fig.~\ref{fig:triggered_dissipation} in the Appendix. Note that although we had chosen $\alpha_{\rm min}=0$
in Eq.~(\ref{eq:alpha_steering_MM}) our sensitive noise triggers causes in the smooth parts of the flow
a small base level of $\alpha\approx 0.03$, see Fig.~\ref{fig:triggered_dissipation}.

\subsection{SPH formulations}
\label{sec:SPH_formulations}
A large variety of SPH formulations  can be found by realizing \citep{price04c}
that the continuity equation can be written  as 
\be
\frac{d\rho}{dt}= \psi \left[ \vec{v}\cdot \nabla \left(
\frac{\rho}{\psi}\right) - \nabla \cdot \left( \frac{\rho \vec{v}}{\psi}\right)\right],
\label{eq:drhodt_psi}
\ee
where $\psi$ is an arbitrary scalar function defined on the particle field,
which can, by straight forwardly applying Eq.~(\ref{eq:nabla_f_a}), be
translated into
\be
\frac{d\rho_a}{dt}=  \sum_b m_b \frac{\psi_a}{\psi_b} \vec{v}_{ab} \cdot
\nabla_a W_h(r_{ab}).
\ee
The consistent momentum equation is given by
\be
\frac{d\vec{v}_a}{dt}= - \sum_b m_b \left( \frac{P_a}{\rho_a^2} \frac{\psi_a}{\psi_b} +
\frac{P_b}{\rho_b^2} \frac{\psi_b}{\psi_a}  \right) \nabla_a W_h(r_{ab})
\label{eq:dvdt_psi}
\ee
and the corresponding equations for the specific internal energy and
the thermokinetic energy $\tilde{e}= u + \vec{v}^2/2$ read
\be
\frac{du_a}{dt}= \frac{P_a}{\rho_a^2} \sum_b m_b \frac{\psi_a}{\psi_b}
\vec{v}_{ab} \cdot \nabla_a W_h(r_{ab})
\label{eq:dudt_psi}
\ee
and
\be
\frac{d \tilde{e}_a}{dt}= - \sum_b \left( \frac{\psi_a}{\psi_b}\frac{P_a
\vec{v}_b}{\rho_a^2} +  \frac{\psi_b}{\psi_a} \frac{P_b
\vec{v}_a}{\rho_b^2} \right) \nabla_a W_h (r_{ab}),
\ee
respectively.

\subsubsection{Version 0 ($V_0$): "traditional SPH" with cubic spline kernels and constant dissipation}
This version serves as a minimum base level: it uses a "vanilla ice" SPH formulation (corresponding to
$\Psi=1$ in the above equations) with a traditional 
cubic spline kernel with 50 neighbor particles, standard kernel gradients and
it applies artificial viscosity with constant parameters $\alpha= \beta/2 = 1$. Explicitly, it reads
\bea
\rho_a &=& \sum_b m_b \bar{W}_{ab} \label{eq:dens_sum_Wab}\\
\left(\frac{d\vec{v}}{dt}\right)_a &=&  - \sum_b m_b \left(\frac{\tilde{P}_a}{\rho_a^2} + \frac{\tilde{P}_b}{\rho_b^2} \right) \nabla_a \bar{W}_{ab}\\
\left(\frac{du}{dt}\right)_a &=& \frac{P_a}{\rho_a^2} \sum_b m_b \vec{v}_{ab} \cdot  \nabla_a \bar{W}_{ab},
\label{eq:V0_set}
\eea
where the $\tilde{P}$s denote pressures that are enhanced by an artificial pressure as discussed in Sec.~\ref{sec:shock_diss}.
Note that we use here always the kernel $\bar{W}_{ab}$, Eq.~(\ref{eq:Wbar}), and its gradient. This is because it is used in the most involved
SPH versions ($V_3, V_4, V_5$) using reproducing kernels and since  a major focus here is on the impact of the gradient accuracy
we want to keep the density calculation the same in all versions. It has to be stressed  that $V_0$ is really a combination of non-optimal options 
and most modern SPH codes use better choices.

\subsubsection{Version 1 ($V_1$): $\Psi=\rho$ with standard kernel gradients}
$V_1$ uses the equation set resulting from $\Psi=\rho$. It has been shown in previous work
\citep{oger07,read10,wadsley17,rosswog20a} that this version is more accurate 
when strong gradients are involved. We show a concise comparison for common choices
of $\Psi$ in Appendix \ref{sec:choice_of_Psi} which clearly favors $\Psi=\rho$. \\
$V_1$ further 
uses standard SPH kernels, but with the Wendland $C4$ kernel \cite{} a much better
kernel function than $V_0$. It  applies shock dissipation as described in Sec.~\ref{sec:shock_diss},
including slope limited reconstruction (using reproducing kernels in the dissipative terms only)
and a steering of the dissipation parameters $\alpha(t)$ and $\beta(t)= 2\alpha(t)$ by means 
of shock and noise triggers, see Eqs.~(\ref{eq:alpha_shock}) and (\ref{eq:alpha_noise}) and we
also apply a small amount of thermal conductivity, see Eq.~(\ref{eq:conduct}) and in the 
"jump terms" of this equation we also use reconstructed quantities.\\
In $V_1$  the density is calculated via Eq.~(\ref{eq:dens_sum_Wab}) and the momentum and energy equation
read
\bea
\left(\frac{d\vec{v}}{dt}\right)_a &=&  \hspace*{-0.3cm}- \sum_b m_b \left(\frac{\tilde{P}_a + \tilde{P}_b}{\rho_a \rho_b} \right) \nabla_a \bar{W}_{ab}\\
\left(\frac{du}{dt}\right)_a &=&\sum_b m_b \left(\frac{\tilde{P}_a}{\rho_a \rho_b} \right) \vec{v}_{ab} \cdot  \nabla_a \bar{W}_{ab} + \left( \frac{du}{dt}\right)_{a,AC},
\label{eq:V1_set}
\eea
where we use  reproducing kernels in the conductivity term, as in the dissipative terms of $V_2$, $V_3$ and $V_5$.\\
$V_0$ and $V_1$ are both fully conservative SPH versions, but  once with bad and once with good choices.

\subsubsection{Version 2 ($V_2$): $\Psi=\rho$ with aLE-gradients}
The only difference between $V_1$ and $V_2$ is that  we now use aLE-gradients, Eq.~(\ref{eq:aLE_gradient}), instead of the
straight forward kernel gradients. We need a replacement of $\nabla_a \bar{W}_{ab}$, Eq.~(\ref{eq:nabla_barW}) and using
Eqs.~(\ref{eq:aLE_gradient}) and (\ref{eq:aLE_replacement}) yields
\bea
2 \nabla^i_a \bar{W}_{ab} &=& \nabla_a^i W_{h_a}(r_{ab}) - \nabla_b^i W_{h_b}(r_{ba})\\
&\rightarrow& C_a^{ik} \nabla_a^k W_{h_a}(r_{ab}) - C_b^{ik} \nabla_b^k W_{h_b}(r_{ba})
\eea
so that the replacement becomes
\be
\nabla^i_a \bar{W}_{ab} \rightarrow \left(\widetilde{\nabla W}\right)^i_{ab} \equiv 
\frac{C_a^{ik} \nabla_a^k W_{h_a}(r_{ab}) + C_b^{ik} \nabla_a^k W_{h_b}(r_{ab})}{2},
\ee
where we have used $\nabla_b^k W_{h_b}(r_{ab})= - \nabla_a^k W_{h_b}(r_{ab})$. 
In  $V_2$ the density is again calculated via Eq.~(\ref{eq:dens_sum_Wab}) and the momentum and energy equations
read
\bea
\left(\frac{d\vec{v}}{dt}\right)_a &=&  \hspace*{-0.3cm}- \sum_b m_b \left(\frac{\tilde{P}_a + \tilde{P}_b}{\rho_a \rho_b} \right) \left(\widetilde{\nabla W}\right)_{ab}\\
\left(\frac{du}{dt}\right)_a &=&\sum_b m_b \left(\frac{\tilde{P}_a}{\rho_a \rho_b} \right) \vec{v}_{ab} \cdot  \left(\widetilde{\nabla W}\right)_{ab} + \left( \frac{du}{dt}\right)_{a,AC}.
\label{eq:V2_set}
\eea

\subsubsection{Version 3 ($V_3$): $\Psi=\rho$ with reproducing kernel gradients}
$V_3$ goes one step further and uses the reproducing kernel gradients, Eq.~(\ref{eq:RPK_gradient})
instead of the  aLE-gradients. In $V_3$  the density is calculated via Eq.~(\ref{eq:dens_sum_Wab}) 
and the momentum and energy equations read
\bea
\left(\frac{d\vec{v}}{dt}\right)_a &=&  \hspace*{-0.3cm}- \sum_b m_b \left(\frac{\tilde{P}_a + \tilde{P}_b}{\rho_a \rho_b} \right) (\nabla \mathcal{W})_{ab}\\
\left(\frac{du}{dt}\right)_a &=&\sum_b m_b \left(\frac{\tilde{P}_a}{\rho_a \rho_b} \right) \vec{v}_{ab} \cdot  (\nabla \mathcal{W})_{ab} + \left( \frac{du}{dt}\right)_{a,AC}.
\label{eq:V3_set}
\eea

\subsubsection{Version 4 ($V_4$):  reproducing kernels and approximate Riemann solver}
The $V_4$ equation set was suggested in \cite{rosswog25a}, it makes use of reproducing kernels and uses
Roe's approximate Riemann solver \citep{roe86}. The starting point is a common  set of Smoothed Particle 
Hydrodynamics (SPH) equations \citep{parshikov02,liu03}
\bea
\rho_a&=&   \sum_b m_b \bar{W}_{ab}\label{eq:dens_sum}\\
\frac{d\vec{v}_a}{dt}&=& - \sum_b m_b \frac{P_a + P_b}{\rho_a \rho_b} \nabla_a \bar{W}_{ab}\label{eq:dvdt1}\\
\frac{du_a}{dt}&=& \frac{1}{2} \sum_b m_b \frac{P_a + P_b}{\rho_a \rho_b} \vec{v}_{ab} \cdot \nabla_a \bar{W}_{ab}.
\label{eq:dudt1}
\eea
The main idea is to solve a one-dimensional Riemann problem at the midpoint between each pair of  
interacting particles $a$ and $b$ and to replace averages of particle values by the solution of a Riemann problem
\be
\frac{P_a + P_b}{2} \rightarrow P_{ab}^\ast \quad {\rm and} \quad  \frac{\tilde{v}_a + \tilde{v}_b}{2} \rightarrow v_{ab}^\ast,
\label{eq:R_subs}
\ee
where the $\ast$ labels the contact discontinuity state in a Riemann
problem,  see e.g. \cite{toro09a}, and  the $ab$-index refers to the
solution between state $a$ and $b$. 
In $V_4$  the density is calculated via Eq.~(\ref{eq:dens_sum_Wab}) and the momentum and energy equations
become
\bea
\frac{d \vec{v}_a}{dt} &=& - \frac{2}{\rho_a} \sum_b V_b P_{ab}^\ast  (\nabla \mathcal{W})_{ab} \label{eq:dvdt3}\\
\frac{d u_a}{dt}&=&  \frac{2}{\rho_a}  \sum_b  V_b  P_{ab}^\ast  (\vec{v}_a - \vec{v}^\ast_{ab}) \cdot (\nabla \mathcal{W})_{ab} \label{eq:dudt3}.
\eea
We use Roe's approximate Riemann solver \citep{roe86} for the star state:
\bea
v_{ab}^\ast &=& \frac{1}{2} \left( (\vec{v}_a + \vec{v}_b) \cdot \hat{e}_{ab} + \frac{P_b - P_a}{C_{\rm RL}}\right)\label{eq:vstar_Roe}\\
P_{ab}^\ast &=& \frac{1}{2} \left( P_a + P_b  +  C_{\rm RL} \frac{ (\vec{v}_b - \vec{v}_a) \cdot \hat{e}_{ab}} {C_{\rm RL}} \right) \label{eq:Pstar_Roe},
\eea
where the "densitized" Roe-averaged Lagrangian sound speed (the dimension is density times velocity) is
\be
C_{\rm RL}= \frac{c_{s,a} \rho_a \sqrt{\rho_a} + c_{s,b} \rho_b \sqrt{\rho_b}}{\sqrt{\rho_a} + \sqrt{\rho_b}}
\ee
 and $c_{\rm s,k}$  is the sound speed of particle $k$.
If, for a moment, we ignore the terms that involve the pressure and velocity differences in Eqs.~(\ref{eq:vstar_Roe}) and 
(\ref{eq:Pstar_Roe}), we have
\be
v_{ab}^\ast \approx \frac{\vec{v}_a + \vec{v}_b}{2} \cdot \hat{e}_{ab}  \quad {\rm and} \quad
P_{ab}^\ast \approx \frac{ P_a + P_b}{2} .
\ee
By inserting these expressions into Eqs.~(\ref{eq:dvdt3}) and (\ref{eq:dudt3}) we obviously recover the inviscid equations
(\ref{eq:dvdt1}) and (\ref{eq:dudt1}), therefore {\em the terms involving the differences in  
Eqs.~(\ref{eq:vstar_Roe}) and (\ref{eq:Pstar_Roe}) are responsible for the dissipation.}
It is in these terms where we apply reconstructed pressures and velocities as explained in Sec.~\ref{sec:shock_diss}.
So, for perfectly reconstructed smooth flows, where the reconstructed values on both sides of the midpoint are the same, the dissipative terms 
vanish and one effectively solves the inviscid hydrodynamics equations.

\subsubsection{Version 5 ($V_5$):  like V4, but with shock dissipation instead of Riemann solver}
$V_5$ solves the Eqs.~(\ref{eq:dens_sum})-(\ref{eq:dudt1}) with the replacement $\nabla_a \bar{W}_{ab} \rightarrow (\nabla \mathcal{W})_{ab}$, but 
instead of using an approximate Riemann solver, it uses shock dissipation. In $V_5$   the momentum and energy equation read
\bea
\frac{d\vec{v}_a}{dt}&=& - \sum_b m_b \frac{\tilde{P}_a + \tilde{P}_b}{\rho_a \rho_b} (\nabla \mathcal{W})_{ab}\\
\frac{du_a}{dt}&=& \frac{1}{2} \sum_b m_b \frac{\tilde{P}_a + \tilde{P}_b}{\rho_a \rho_b} \vec{v}_{ab} \cdot  (\nabla \mathcal{W})_{ab}
+  \left( \frac{du}{dt}\right)_{a,AC}.
\eea
A comparison between the results obtained with $V_4$ and $V_5$ therefore allows to gauge of the effect 
of shock dissipation vs approximate Roe Riemann solver.\\
In all of the shown tests we use the Wendland C4 kernel \citep{wendland95}
\be
W_{C^4}(r)= \sigma(1-q)^6_{+} \; \left(1 + 6 q + \frac{35}{3}q^2\right),
\ee
where $q= r/2h$ so that the kernel support extends to $2h$ and the corresponding normalization is $\sigma= \frac{495}{256 \; \pi}$. 
We use exactly 250 contributing neighbors within the support size 2$h_a$ of each particle $a$. How exactly this is achieved by 
means of our recursive coordinate bisection tree \citep{gafton11} is described in  detail in \cite{rosswog20a} .

\section{Results}
\label{sec:results}

\subsection{Questions to be addressed}
In this section we perform a number of common benchmark tests to probe the abilities
of the different SPH formulations. In particular, we are interested in the following 
\bi
\i Which quantity should be chosen for the scalar $\Psi$? 
\i Is an additional  trigger on noise beneficial in complex shock tests? 
\i How important is the gradient accuracy for specific tests?
\i How much better are the reproducing kernel gradients than the standard kernel gradients 
   and the  aLE gradients? The aLE gradients are substantially cheaper to calculate (only a $3 \times 3$ 
   matrix needs to be inverted) than the RPK gradients (much more matrix algebra, see Appendix
   \ref{sec:App_RPK}), and, from a pragmatic point of view, one may wonder whether the aLE-gradients
   are good enough.
\i How does the recently suggested RPK gradient hydrodynamics version that uses Roe's Riemann solver
   ($V_4$) compare to the same equation set, but with shock dissipation rather than a Riemann solver?  
   In other words, how different are the results between Riemann solver and shock dissipation? 
\ei
To keep the parameter space to be explored under controle, we address the first two questions in Appendices
\ref{sec:choice_of_Psi} and \ref{sec:effectiveness_noise_trigger}). We show in Appendix \ref{sec:choice_of_Psi}
several tests for common choices of $\Psi$ and find a clear preference for $\Psi= \rho$. This result is consistent
with earlier work \citep{oger07,read10,wadsley17,rosswog20a}. In Appendix \ref{sec:effectiveness_noise_trigger},
we compare for a Schulz-Rinne test case dissipation triggered only with the \cite{cullen10} shock trigger with
a simulation that additionally applies the noise trigger described above. As shown in Figs.~\ref{fig:effect_noise_trigger} 
and \ref{fig:triggered_dissipation}, this trigger provides an appropriate amount of dissipation without over-triggering 
dissipation in smooth flows.\\
The other questions are addressed in a set of challenging  benchmarks, all of which are preformed with the
3D code. The first three cases are shock tests (Secs~\ref{subsec:Sedov} to \ref{subsec:Riemann2}), followed
by two instability tests and, finally, a selection of challenging Schulz-Rinne Riemann problems.

\begin{figure*}
\includegraphics[width=6in]{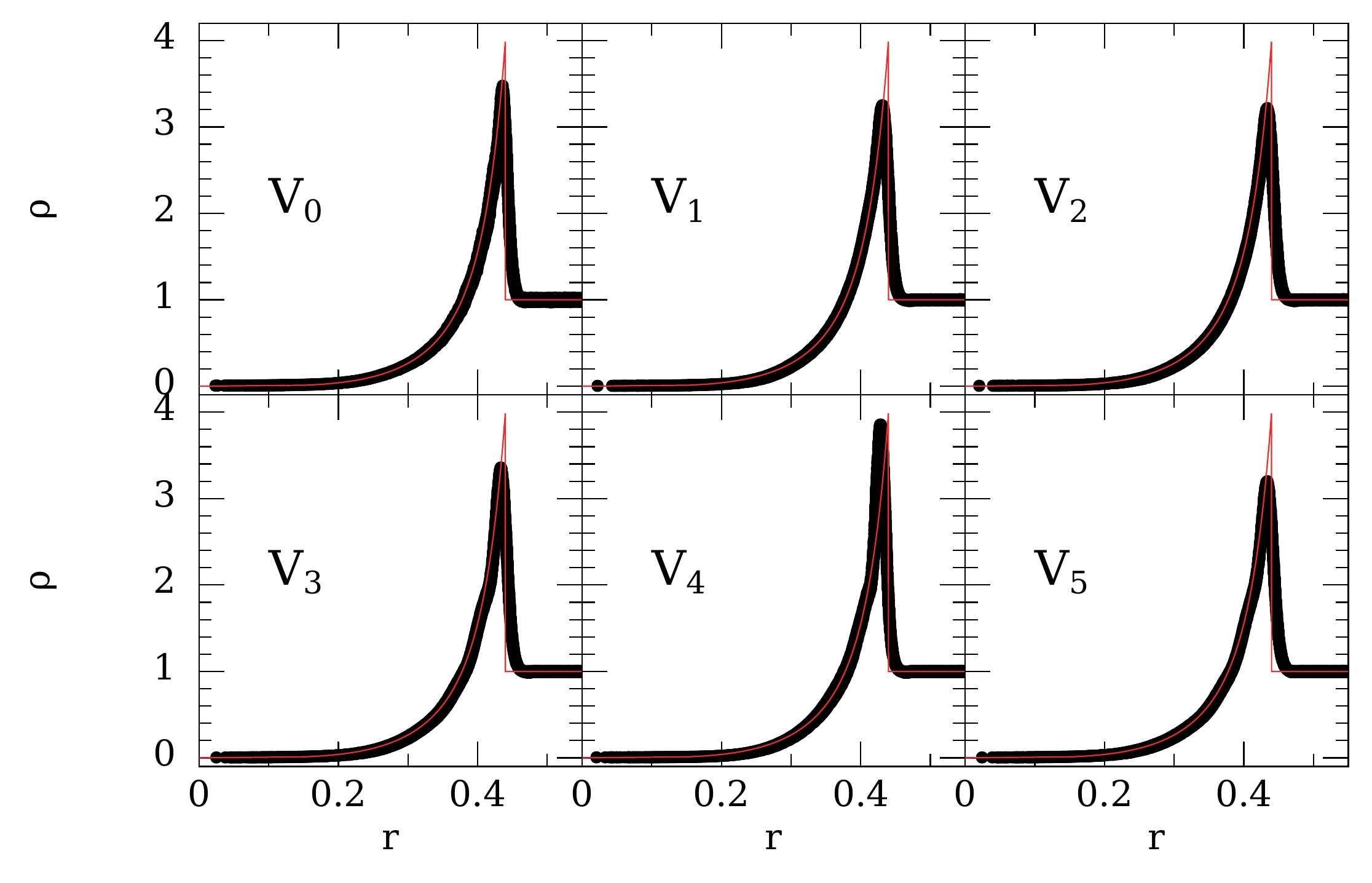} 
\caption{Density in a Sedov explosion at $t=0.9$ for all SPH versions.}
\label{fig:Sedov_comp}
\end{figure*} 
\subsection{3D Sedov explosion}
\label{subsec:Sedov}
The Sedov--Taylor test, a strong initial point-like explosion
expanding into a low-density environment, is one of the standard
benchmark tests for hydrodynamics codes \citep{sedov59,taylor50}.
For an explosion energy $E$ and a density of the ambient medium
$\rho$, the blast wave propagates to the radius
$r(t)= \beta (E t^2/\rho)^{1/5}$ at  time $t$, where $\beta$ depends on the
adiabatic exponent of the gas ($\approx 1.15$ in 3D  for the
$\Gamma=5/3$ we use).  In the strong explosion limit, the density
jumps by a factor of  $\rho_2/\rho_1= (\Gamma + 1)/(\Gamma-1)= 4$ in
the shock front, where the numerical value refers to our chosen 
value of $\Gamma$. We place $128^3$ particles according to a
Centroidal Voronoi Tessellation \citep{du99}  in the volume
[-0.5,0.5]$^3$. For an even better particle distribution we perform
additional sweeps according to our ``Artificial Pressure Method'' (APM),
see  \citet{rosswog20a} for a detailed description.\\
Here all  the different versions perform pretty well, which is in part due to the
careful preparation of the initial particle distribution. From the shock dissipation
versions, the reference version $V_0$ reaches the highest peak, because 
the chosen cubic spline kernel with only 50 neighbors yields the smallest
kernel support size. The Riemann solver version $V_4$ reaches clearly the 
highest peak, and in the wake of the shock the RPK gradient versions ($V_3..V_5$)
show a small tendency for oscillations. Among the RPK versions $V_3$ reaches the largest 
peak height.

\begin{figure*} 
   \includegraphics[width=5in]{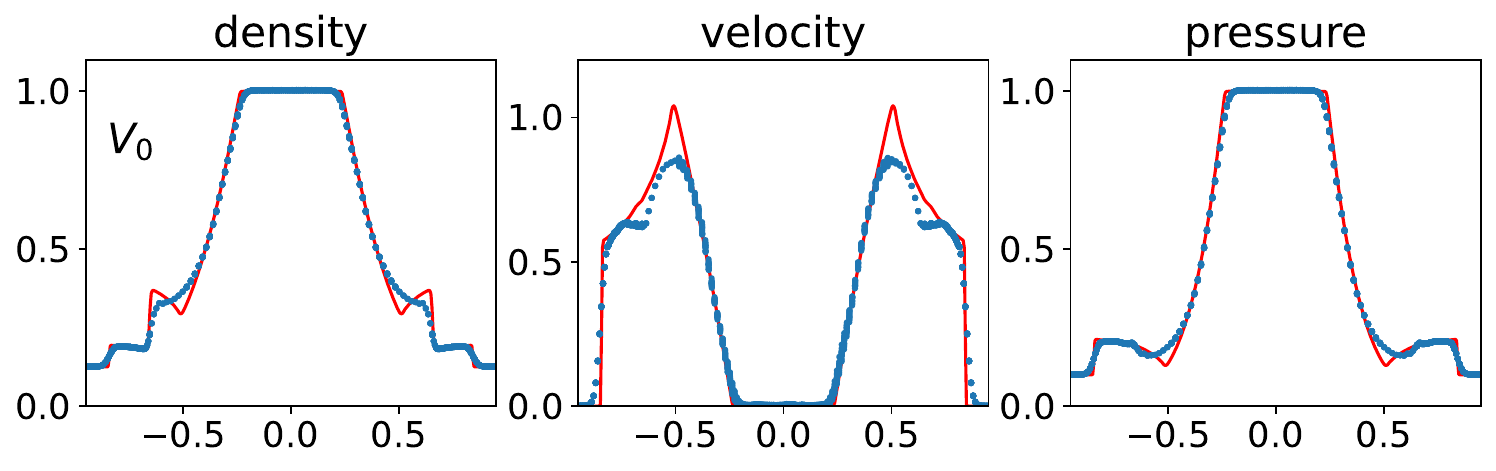} \\
   \includegraphics[width=5in]{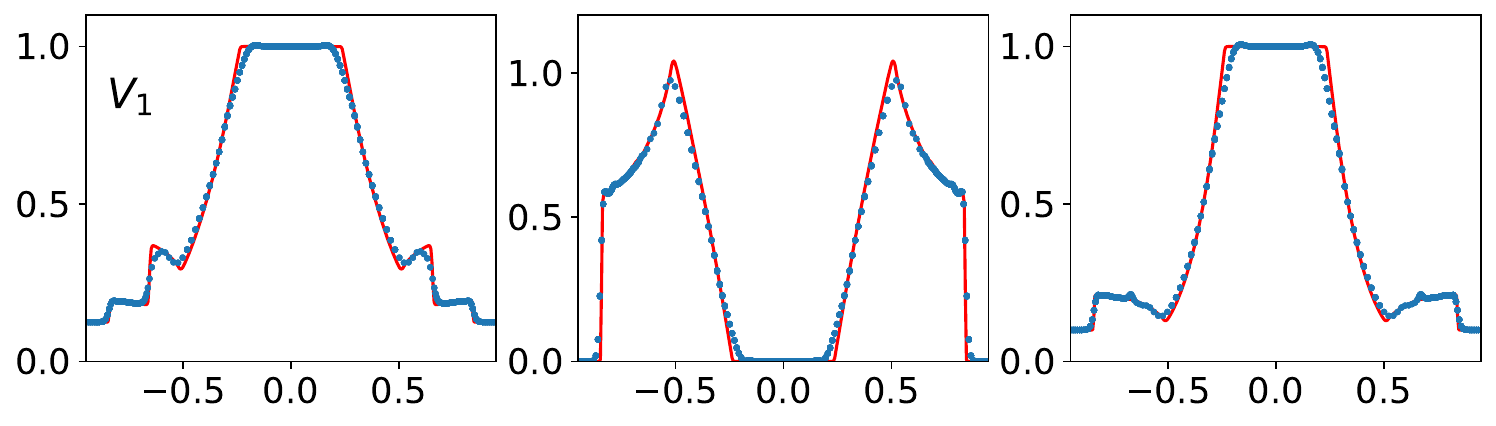} \\
   \includegraphics[width=5in]{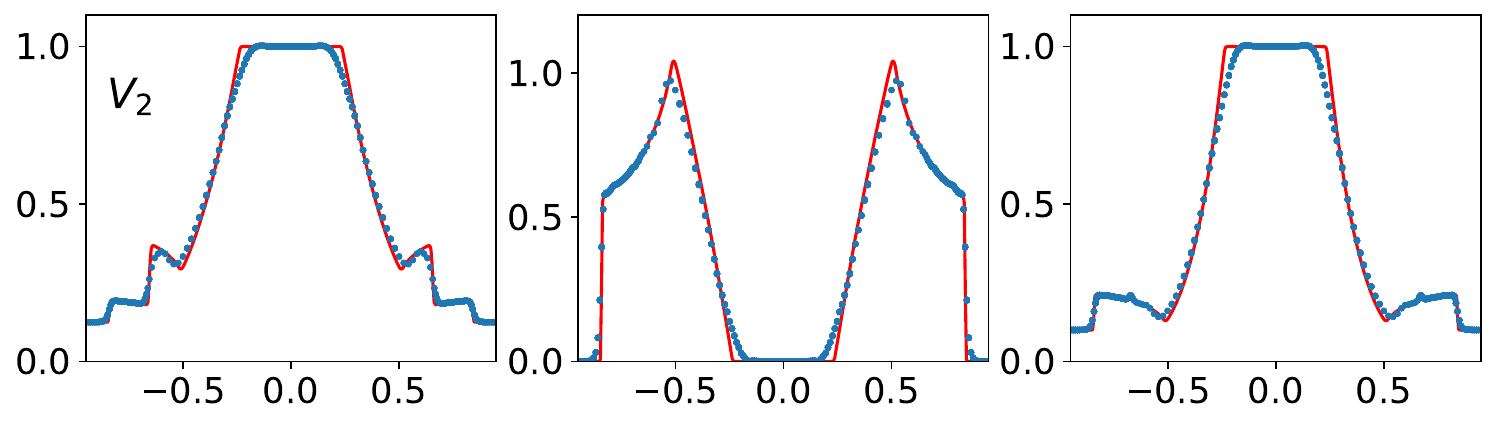} \\
   \includegraphics[width=5in]{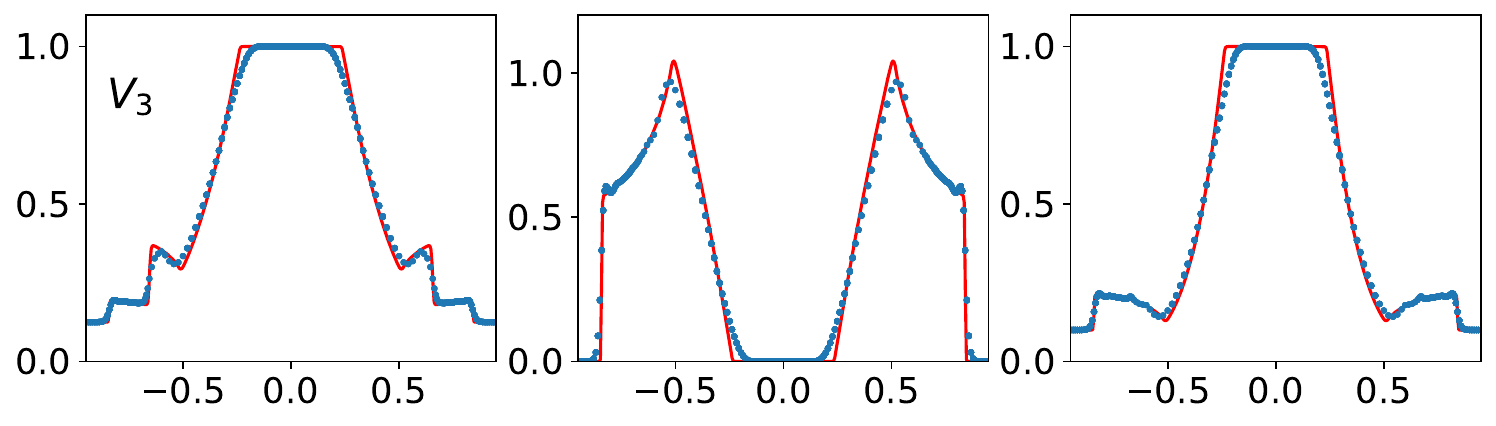} \\
   \includegraphics[width=5in]{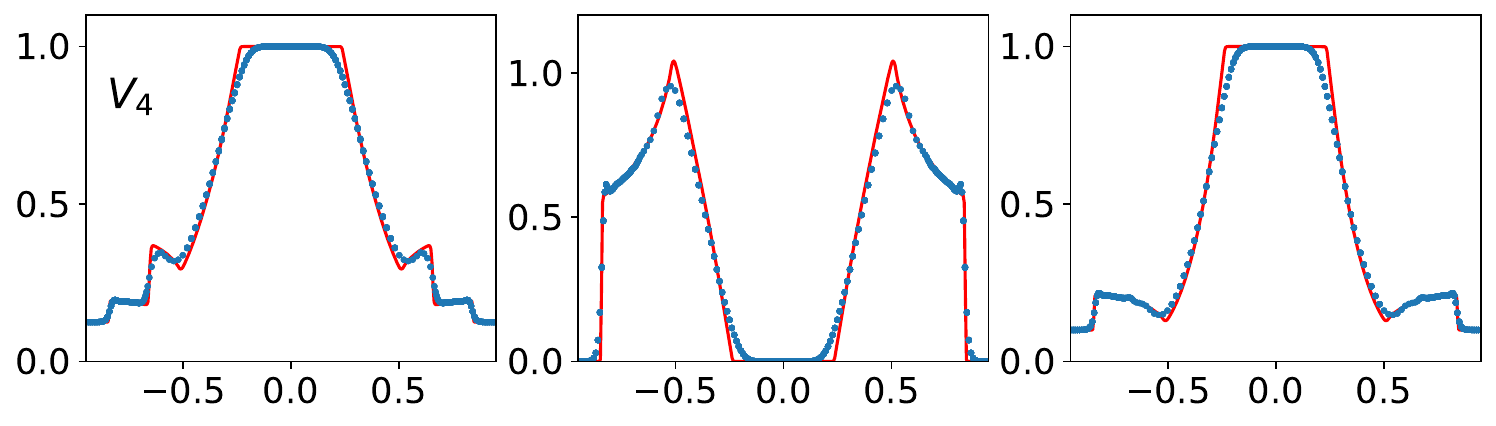} \\
   \includegraphics[width=5in]{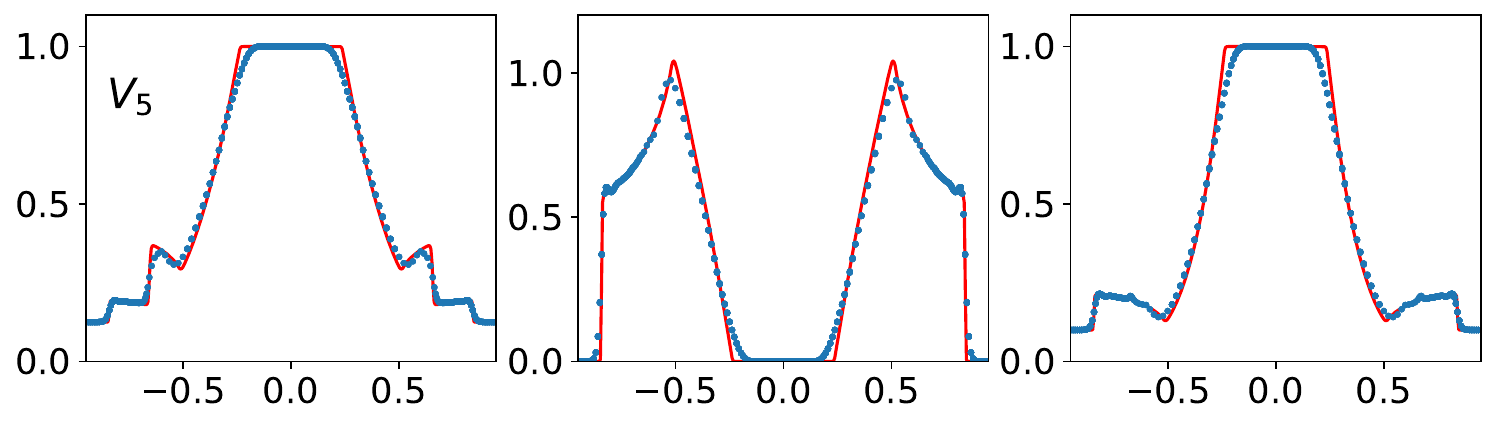} \\      
   \caption{3D Riemann problem 1 for all SPH versions at $t=0.2$}
   \label{fig:Riemann1}
\end{figure*}
\subsection{3D spherical Riemann Problem 1}
\label{subsec:Riemann1}
We choose the same parameters as \cite{toro09a}  (apart from a shift of the origin): 
the computational domain is $[-1,1]^3$ and the initial conditions are chosen as:
\be
(\rho,\vec{v},P)=
 \left\{
\begin{array}{l}
(1.000,0,0,0,1.0) \quad {\rm for \; \; r < 0.5}\\
(0.125,0,0,0,0.1) \quad {\rm else.}
\end{array}
\right.
\ee
The solution exhibits a spherical shock wave, a spherical contact surface traveling in the same direction, 
and a spherical rarefaction wave traveling toward the origin. As initial conditions, we simply placed  $200^3$ 
particles on a cubic lattice within $[-1,1]^3$, together with the surrounding "frozen" particles as the boundary condition. \\
In Fig.~\ref{fig:Riemann1} we show our particle results in a strip around the $x-$axis ($|y| < 0.018, |z| < 0.018$) 
compared with a $400^3$ grid cell calculation with the Eulerian weighted average flux method \citep{toro09a}. \\
Overall, we find very good agreement with the reference solution for all versions $V_1$..$V_5$.
Clearly worst is $V_0$ which smears out the contact discontinuity region in the density and produces a spurious 
dip in the velocity. Simply changing to better SPH choices, $V_1$, substantially improves the agreement with the 
reference solution. There is only a very small density overshoot at the beginning of the  edge of the 
rarefaction wave ($x\approx 0.2$), a tiny overshoot of the velocity at the shock front and a very small "pressure blip" at the contact 
discontinuity.  Using the aLE-approximation, $V_2$, slightly improves on the pressure blip and essentially shows 
no velocity overshoot at the shock. $V_3$ with the RPK gradients improves on the pressure blip, but also shows 
some velocity overshoot at the shock. $V_4$ with the RPK Riemann solver approach essentially gets rid of the 
pressure blip, also shows the velocity overshoot and slightly larger rarefaction wave. This is because the dissipation 
from the Riemann solver is applied both in expanding and compressing flows, while the shock dissipation of the other 
versions is only applied in compressing flows, see Eq.~(\ref{eq:mu_vis}). Overall, all modern versions perform very 
well in this test, only the baseline version $V_0$ shows deficiencies.

\begin{figure*} 
   \includegraphics[width=5in]{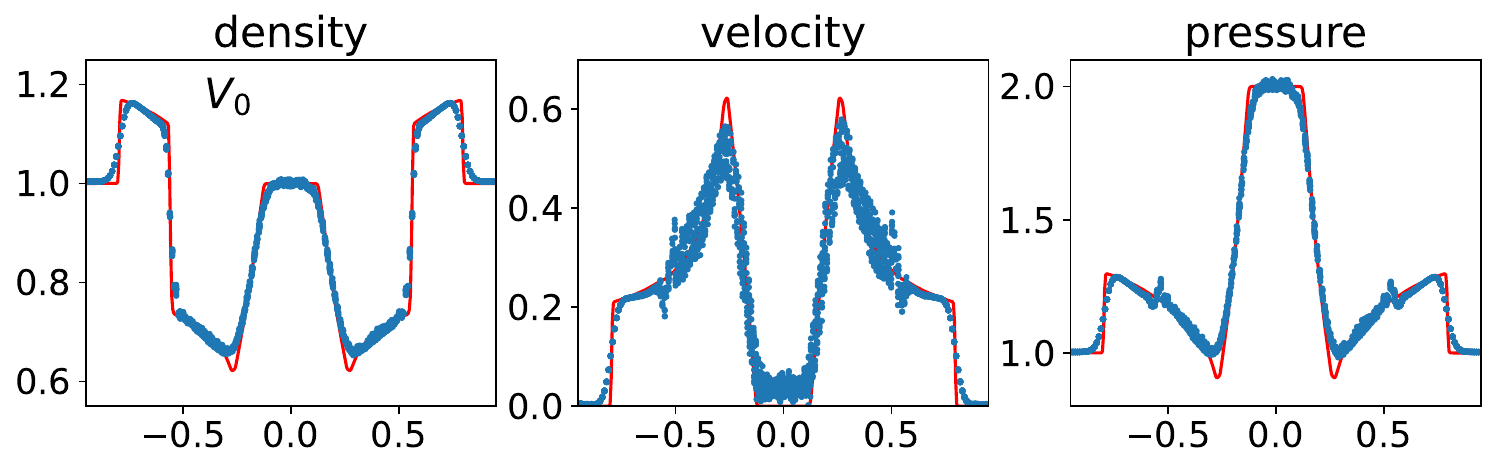} \\
   \includegraphics[width=5in]{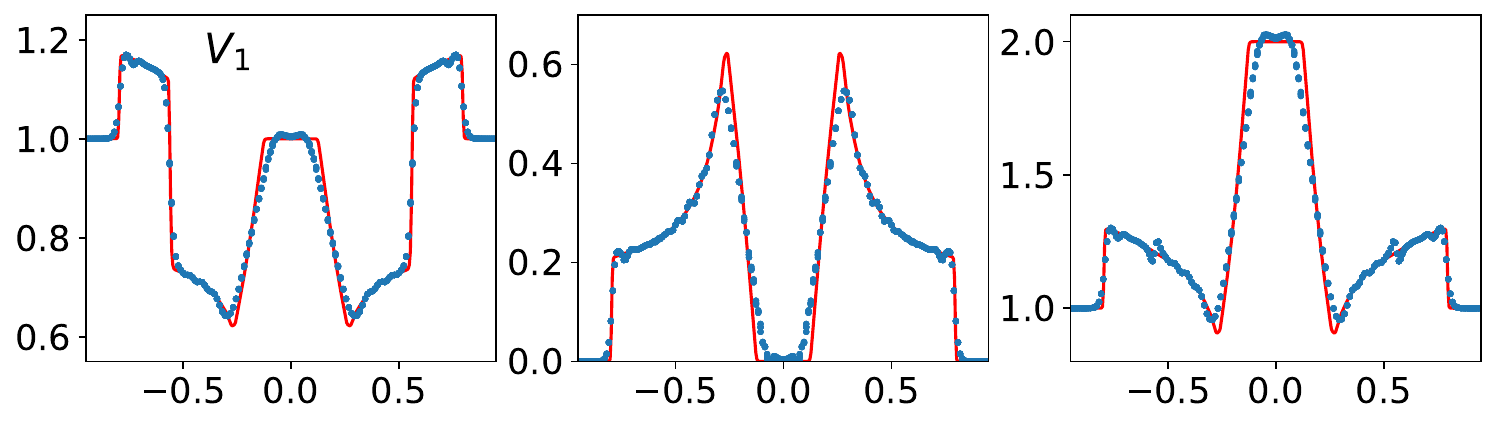} \\
   \includegraphics[width=5in]{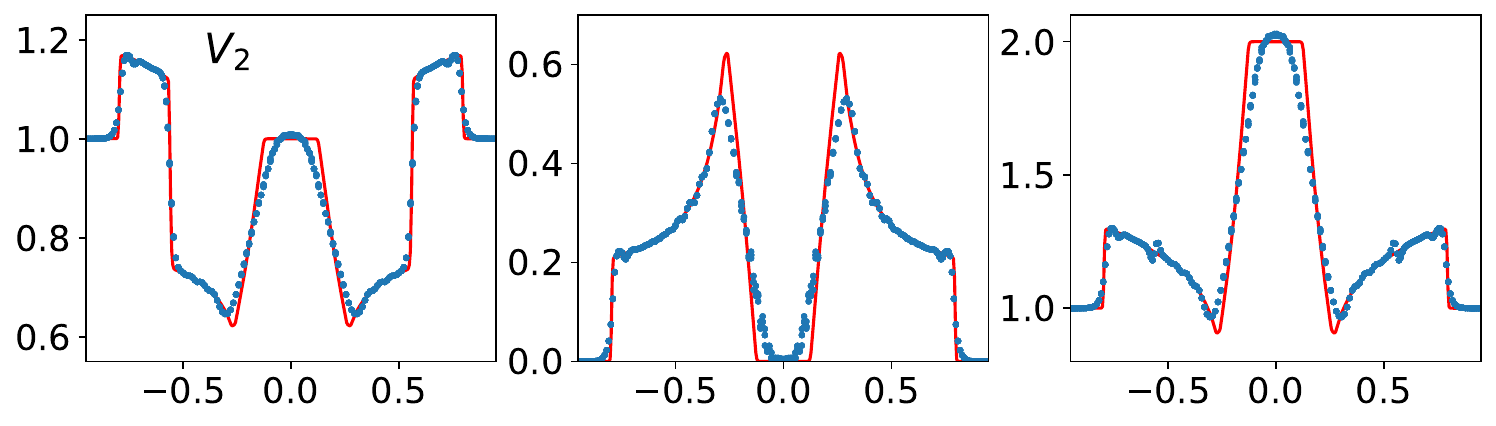} \\
   \includegraphics[width=5in]{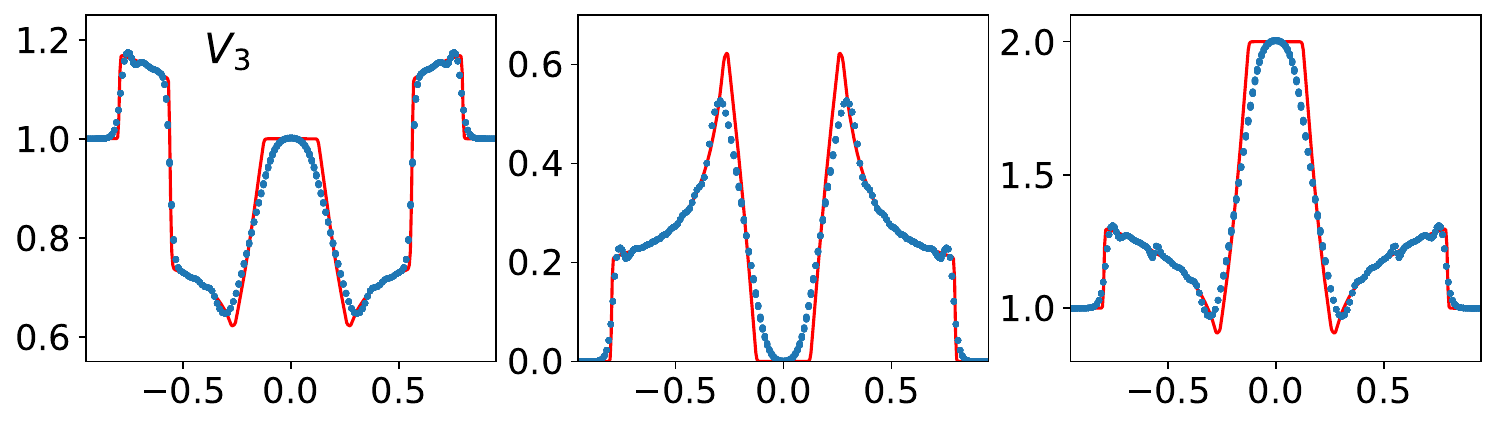} \\
   \includegraphics[width=5in]{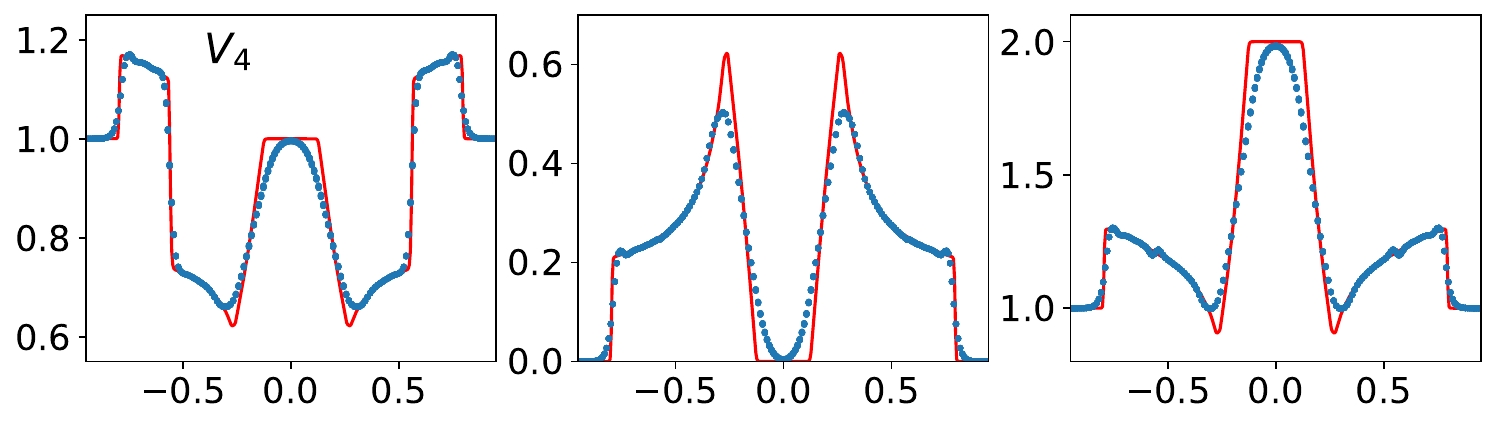} \\
   \includegraphics[width=5in]{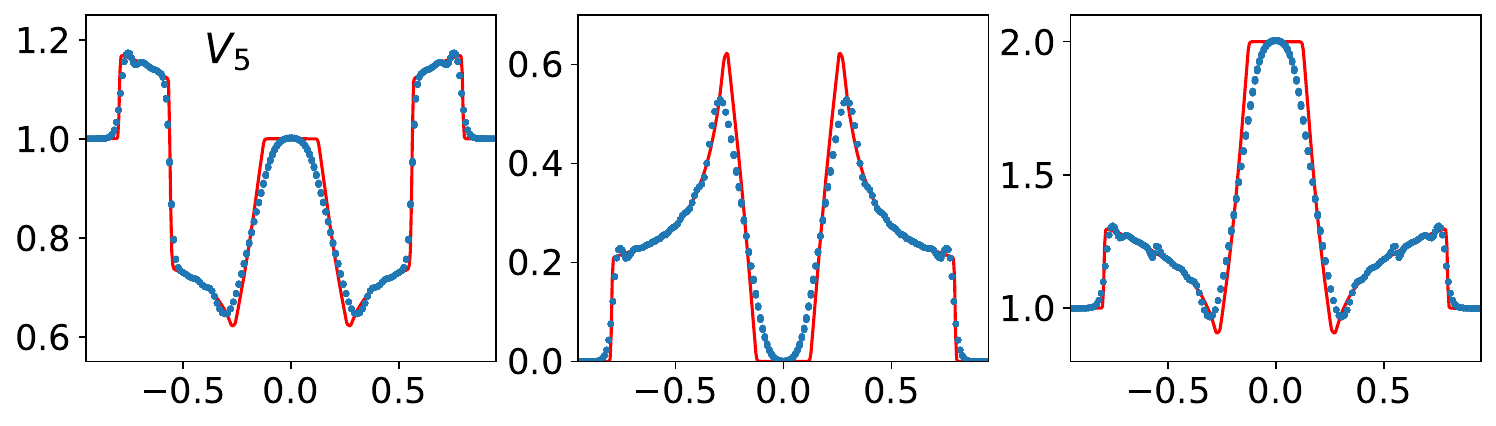} \\      
   \caption{3D Riemann problem 2 for all SPH versions at $t=0.2$}
   \label{fig:Riemann2}
\end{figure*}
\subsection{3D spherical Riemann Problem 2}
\label{subsec:Riemann2}
In a second spherical blast wave problem \citep{toro09a} we start from  
\be
(\rho,\vec{v},P)=
 \left\{
\begin{array}{l}
(1.0,0,0,0,2.0) \quad {\rm for \; \; r< 0.5}\\
(1.0,0,0,0,1.0) \quad {\rm else}
\end{array}
\right.
\ee
and place again $200^3$ particles in the same straightforward way as
in the first blast wave problem.
\\
We show in Fig.~\ref{fig:Riemann2} the results for all five versions. Again, the baseline version $V_0$ shows
substantial artifacts in the velocity. The Riemann solver RPK version $V_4$ shows the smallest oscillations in
the solution, but it is more dissipative at the beginning of the rarefaction wave, again because it applies dissipation
to both converging and expanding flows. Here we do see a difference of the gradient accuracy: both the standard 
SPH gradient ($V_1$) and the improved aLE-gradient ($V_2$) show more velocity noise in the inner regions than
the RPK gradient versions $V_3 .. V_5$.

\begin{figure*} 
   \includegraphics[width=6in]{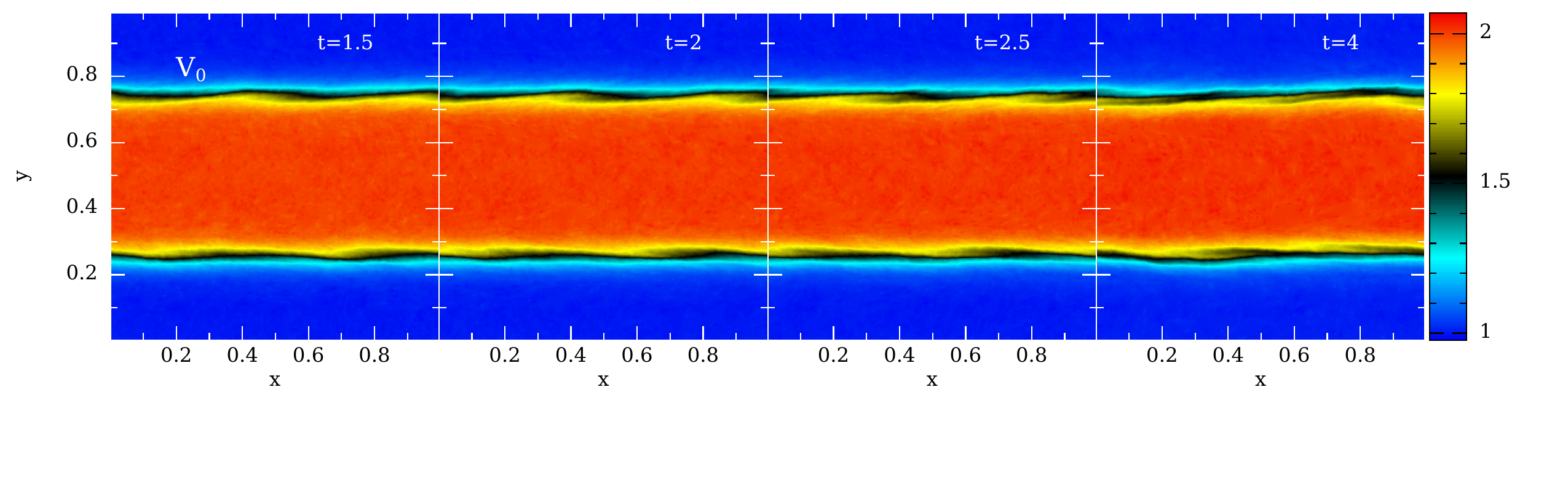} \\
   \vspace*{-1cm}
   \includegraphics[width=6in]{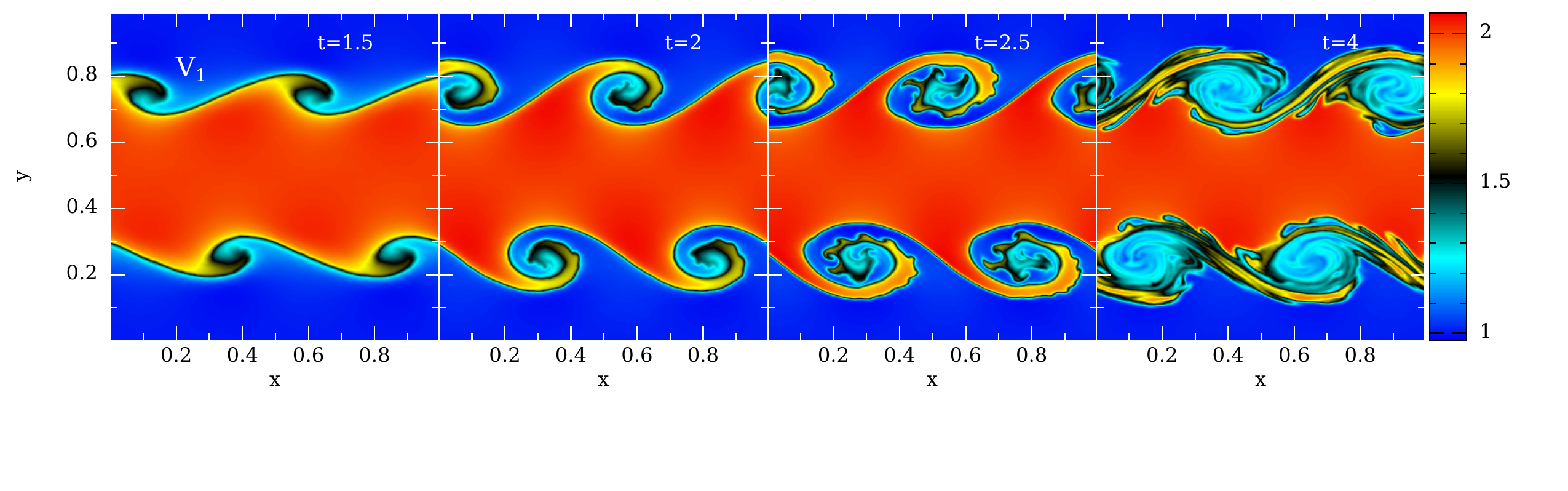} \\
   \vspace*{-1cm}
    \includegraphics[width=6in]{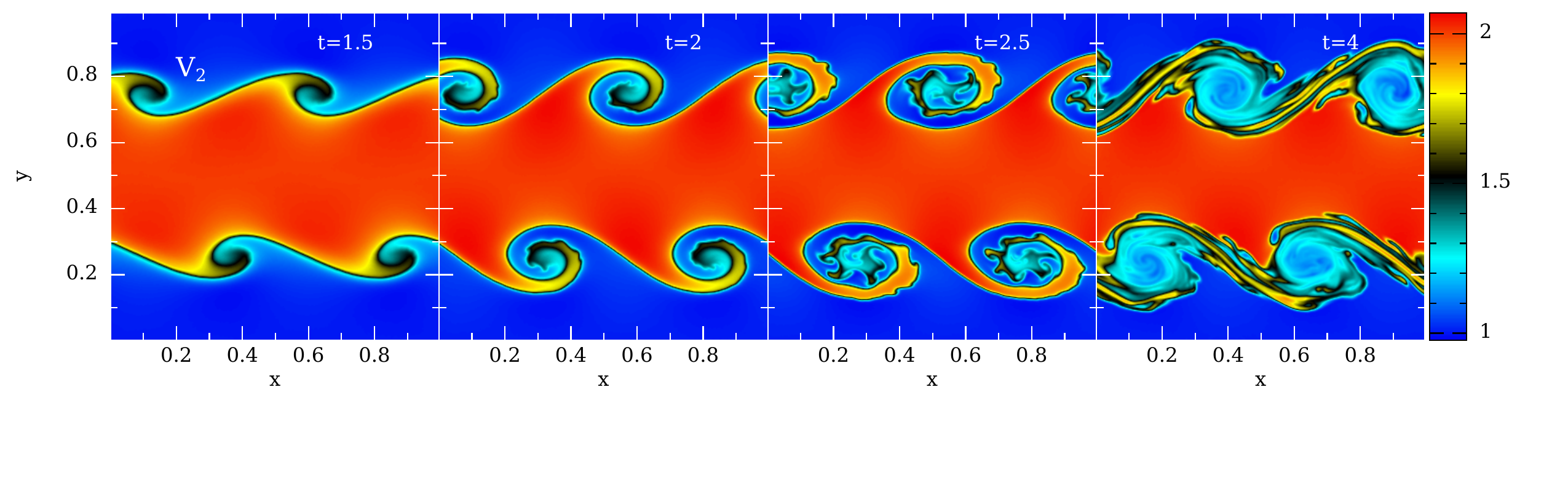}\\
    \vspace*{-1cm}
    \includegraphics[width=6in]{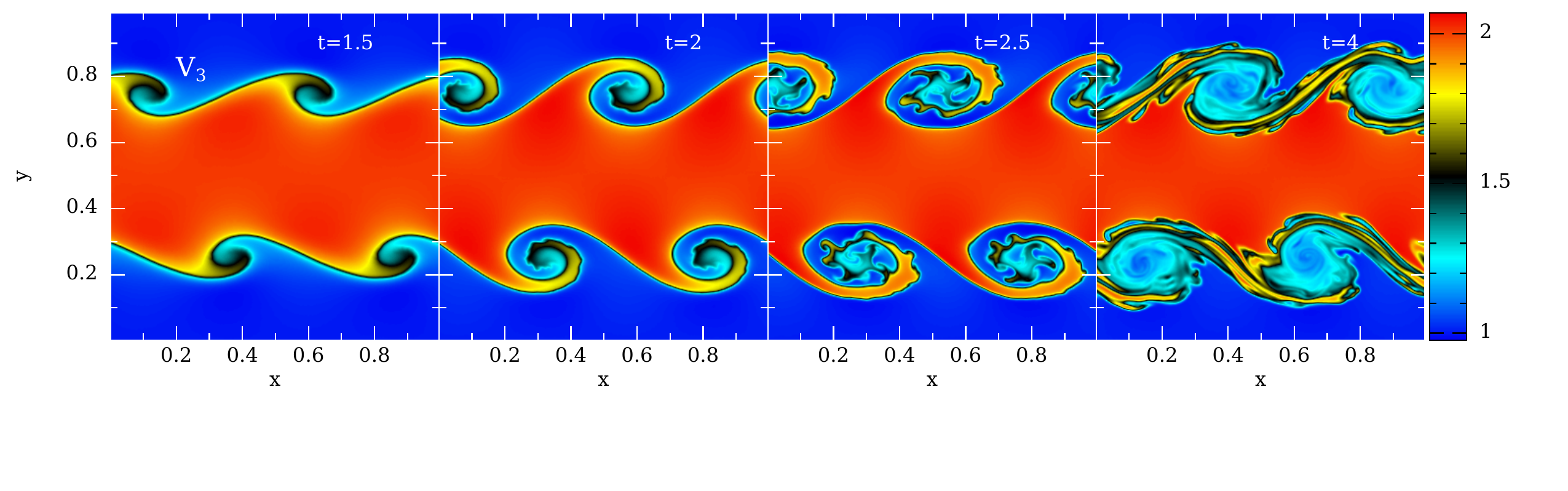}\\ 
    \vspace*{-1cm}
    \includegraphics[width=6in]{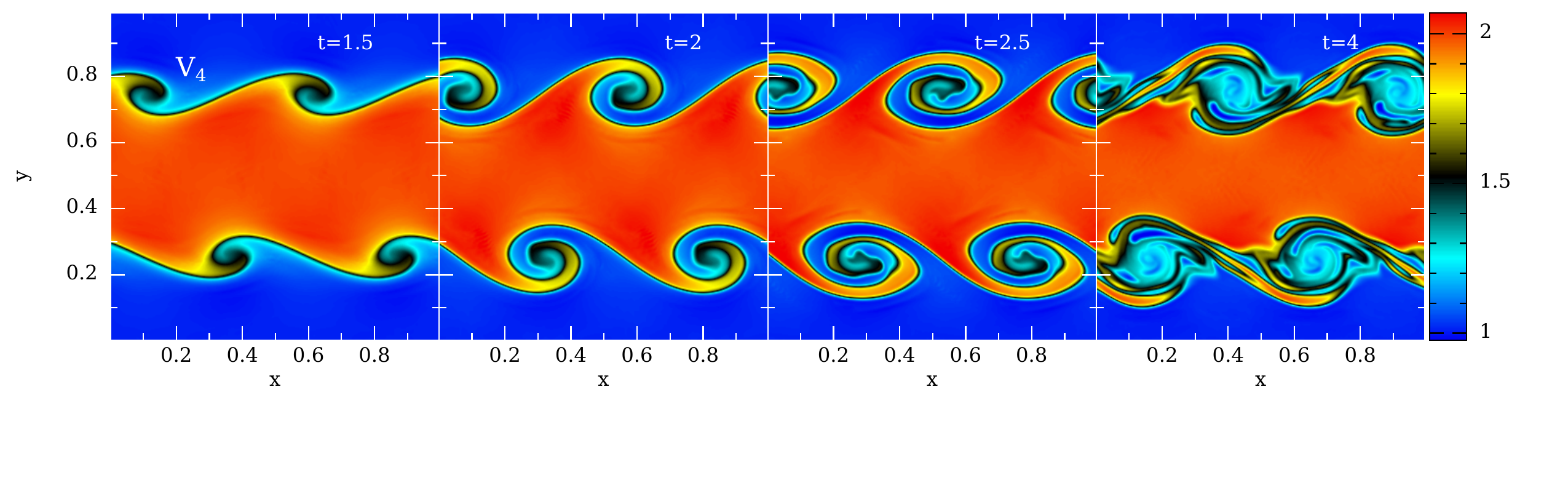}\\
    \vspace*{-1cm}  
    \includegraphics[width=6in]{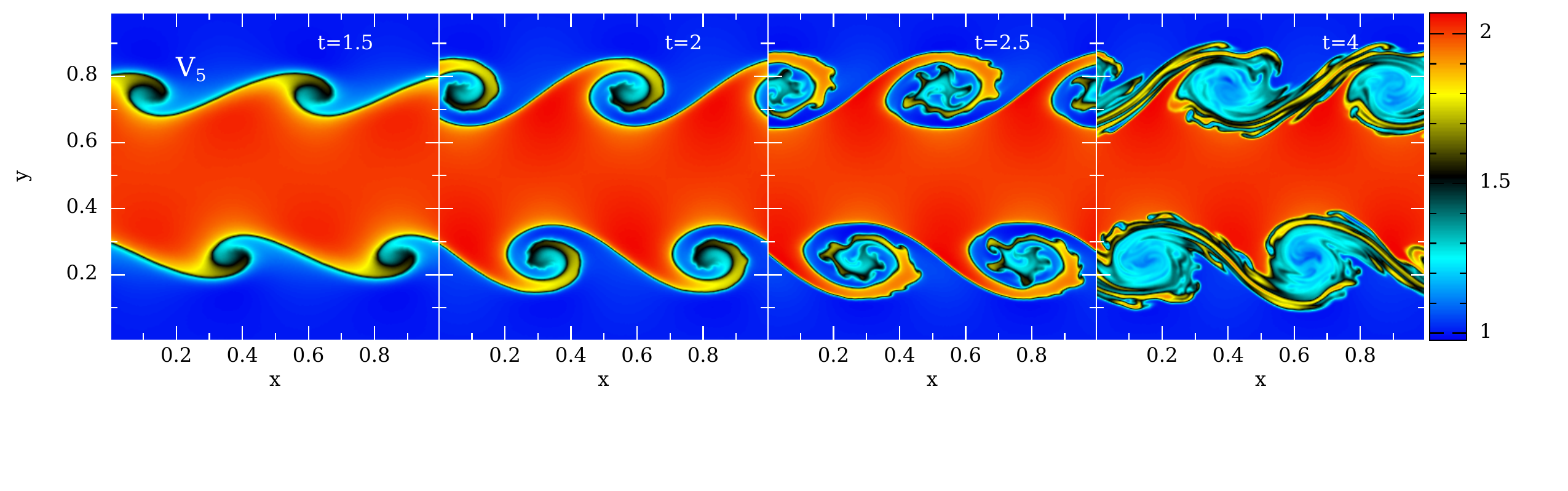}\\
     \vspace*{-1cm}  
    \caption{Kelvin-Helmholtz instability for all SPH versions at $512^2$.}
   \label{fig:KeHe512}
\end{figure*}
%
\begin{figure} 
   \includegraphics[width=3.5in]{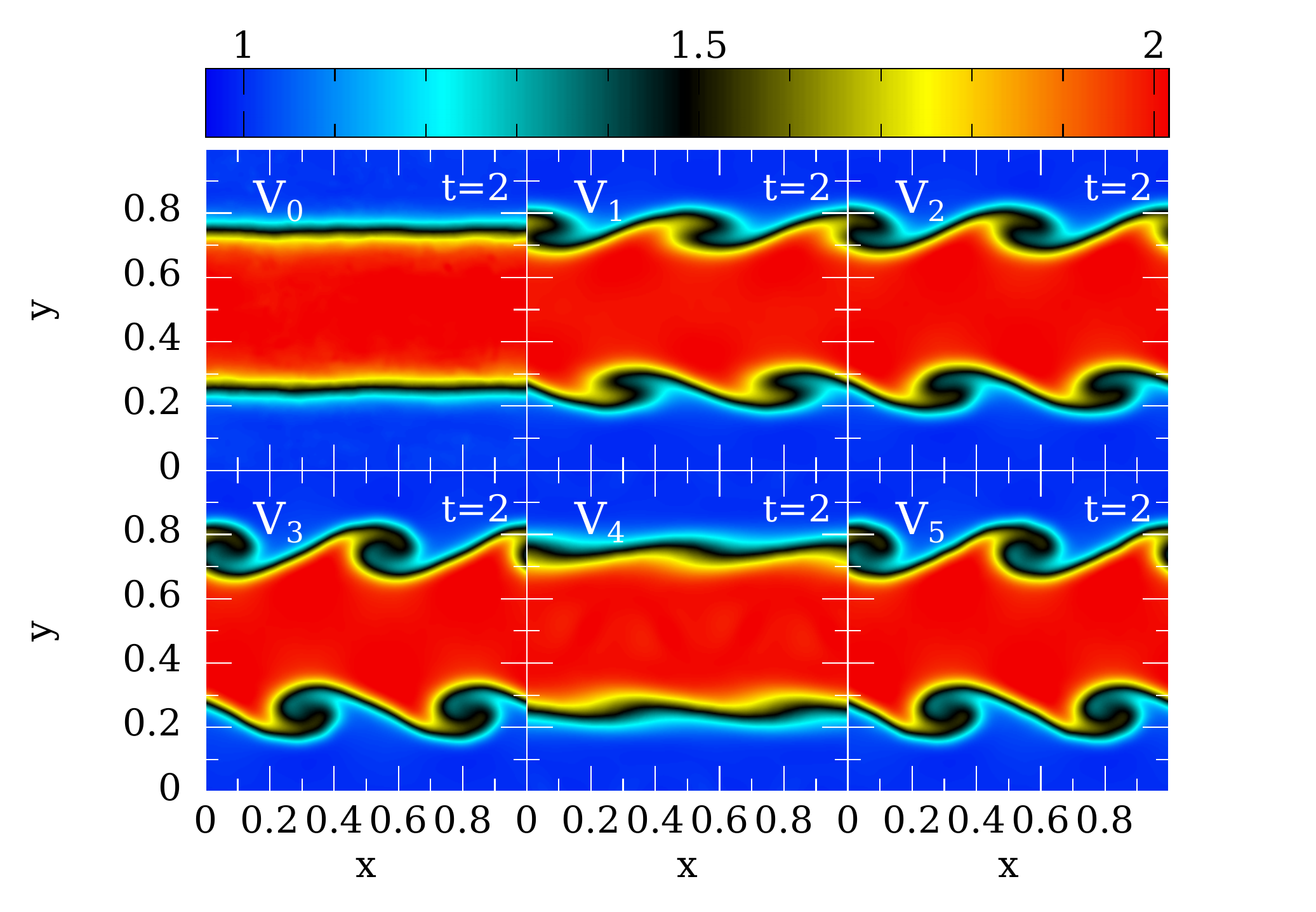} \\
   \vspace*{-0.7cm} 
   \caption{Density of Kelvin-Helmholtz instability at low resolution ($64^2$) for all versions.} 
   \label{fig:KeHe64}
\end{figure}
\begin{figure*} 
   \centerline{
   \includegraphics[width=2.8in]{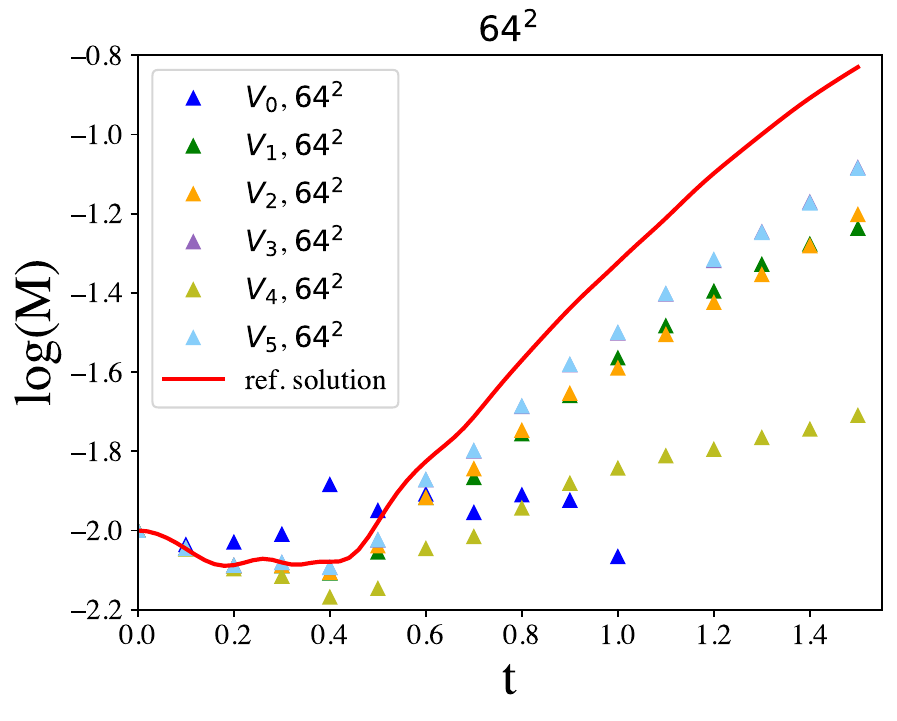} 
   \includegraphics[width=2.8in]{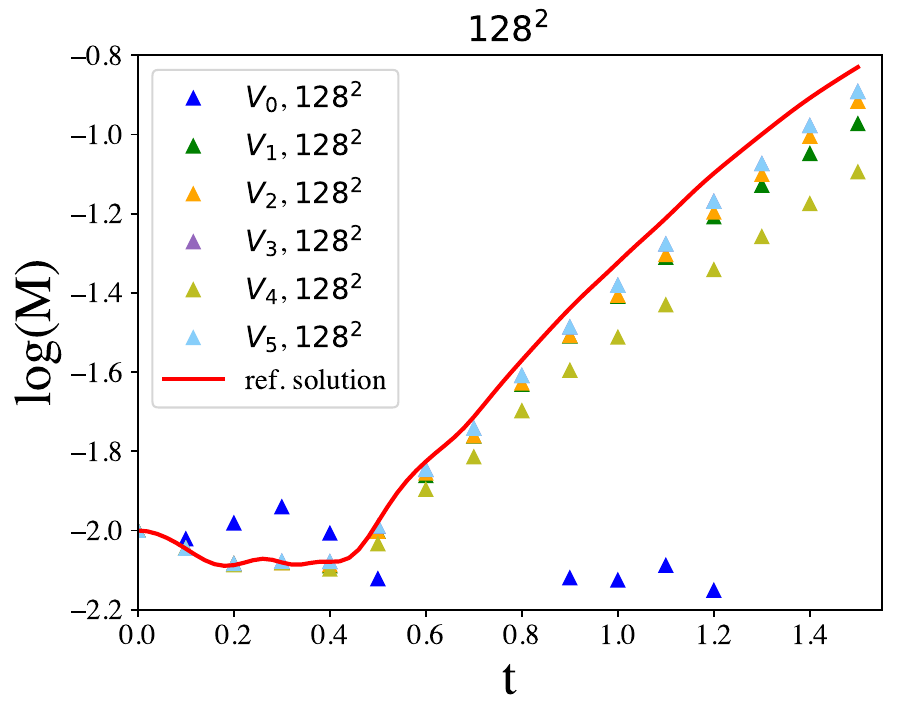} }
   \centerline{
   \includegraphics[width=2.8in]{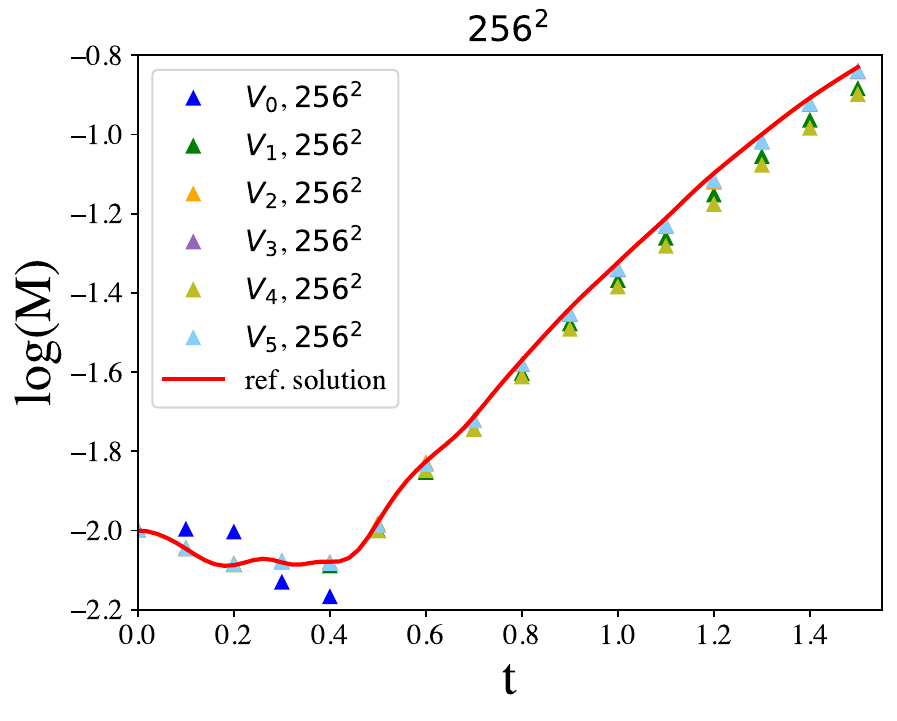} 
   \includegraphics[width=2.8in]{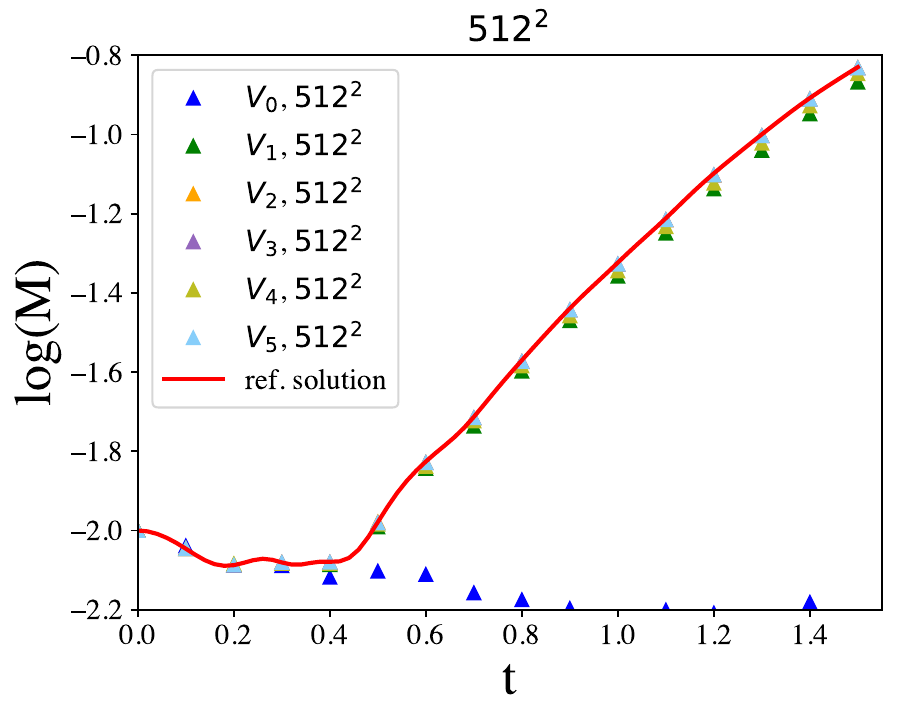} 
   }
   \caption{Mode growth at $64^2$, $128^2$, $256^2$ and $512^2$  compared to 
   a very high resolution simulation ($4096^2$) obtained with the PENCIL code. Note that
   $V_0$ (dark blue triangles) never grows properly, not even at $512^2$ and several of
   the $V_0$ symbols are below the shown scale. The versions $V_3$ and $V_5$ perform nearly
   identically and their symbols lie on top of each other so that only the symbols for $V_5$ are visible.}
   \label{fig:mode_growth}
\end{figure*}

\subsection{Kelvin-Helmholtz}
\label{sec:KeHe}
Traditional versions of SPH have been shown to struggle with weakly triggered Kelvin-Helmholtz (KH) instabilities \citep{agertz07,mcnally12}, 
although many recent studies with more sophisticated numerical methods yielded very good results \citep{frontiere17,rosswog20a,sandnes25}.
We focus here on a test setup in which traditional SPH has been shown to fail, even at a rather high resolution  in 2D, see \cite{mcnally12}.
We follow the latter paper (similar setups were used in \cite{frontiere17}, \cite{rosswog20a} and \cite{sandnes25}), but 
we use the full 3D code and set up the "2D" test as a thin 3D slice with $N \times N \times 20$ particles (referred to as $"N^2"$).
For simplicity, the particles are initially placed on a cubic lattice. Periodic boundary conditions are enforced by placing 
appropriate particle copies outside of the "core" volume. The test is initialized as:
\be
\rho(y)=
 \left\{
\begin{array}{l}
\rho_1 - \rho_m e^{(y - 0.25)/\Delta} \quad {\rm for \; \; 0.00 \le y < 0.25}\\
\rho_2 + \rho_m e^{(0.25 - y)/\Delta} \quad {\rm for \; \; 0.25 \le y < 0.50}\\
\rho_2 + \rho_m e^{(y - 0.75)/\Delta} \quad {\rm for \; \; 0.50 \le y < 0.75}\\
\rho_1 - \rho_m e^{(0.75 - y)/\Delta} \quad {\rm for \; \; 0.75 \le y < 1.00},\\
\end{array}
\right.
\ee
where $\rho_1= 1$, $\rho_2= 2$, $\rho_m= (\rho_1 - \rho_2)/2$ and $\Delta= 0.025$.
The velocity is set up as
\be
v_x(y)=
 \left\{
\begin{array}{l}
v_1 - v_m e^{(y - 0.25)/\Delta} \quad {\rm for \; \; 0.00 \le y < 0.25}\\
v_2 + v_m e^{(0.25 - y)/\Delta} \quad {\rm for \; \; 0.25 \le y < 0.50}\\
v_2 + v_m e^{(y - 0.75)/\Delta} \quad {\rm for \; \; 0.50 \le y < 0.75}\\
v_1 - v_m e^{(0.75 - y)/\Delta} \quad {\rm for \; \; 0.75 \le y < 1.00}\\
\end{array}
\right.
\ee
with $v_1$= 0.5, $v_2= -0.5$, $v_m= (v_1-v_2)/2$ and a small velocity perturbation in 
$y$-direction is introduced as $v_y=  0.01 \sin(2\pi x/\lambda)$ with the perturbation 
wave length $\lambda= 0.5$. In the linear regime, a Kelvin-Helmholtz instability grows 
in the incompressible limit on a characteristic time scale of
\be
\tau_{\rm KH}= \frac{(\rho_1+\rho_2) \lambda}{\sqrt{\rho_1 \rho_2} |v_1-v_2|},
\ee
with $\tau_{\rm KH}\approx 1.06$ for the chosen parameters.
For our tests, we chose again a polytropic equation of state with
exponent $\Gamma=5/3$. \\
We show in Fig.~\ref{fig:KeHe512} a $512^2$ realization of this test for all SPH versions.
As in \cite{mcnally12}, the traditional SPH version does not show any sign of growth
until the end of the simulation ($t=10$). All other versions show a healthy growth of the 
instability. $V_4$ is slightly lagging behind in the growth and shows an enormous degree 
of symmetry. Its Kelvin-Helmholtz billows also do not break up into smaller scale substructures.
As an additional experiment, we ran a low resolution version ($64^2$) of this test, see 
Fig.~\ref{fig:KeHe64}, which confirms this picture: $V_0$ does not grow 
all, the Riemann solver version $V_4$ is seriously delayed, the standard gradient version ($V_1$)
and the aLE-gradient version ($V_2$) grow healthily, but slightly slower than the RPK versions
$V_3$ and $V_5$.\\
To quantify these results further, we measure the growth rates (calculated exactly as in \cite{mcnally12})
for all versions at $64^2$, $128^2$, $256^2$ and $512^2$, see Fig.~\ref{fig:mode_growth}.
These results are compared with a high resolution ($4096^2$) simulation with the PENCIL 
code \citep{brandenburg02}. This plot confirms the visual impression that the Riemann solver 
version $V_4$ grows substantially slower than even the standard SPH kernel gradient version $V_1$.

\begin{figure*} 
   \hspace*{-0.5cm}\includegraphics[width=20cm]{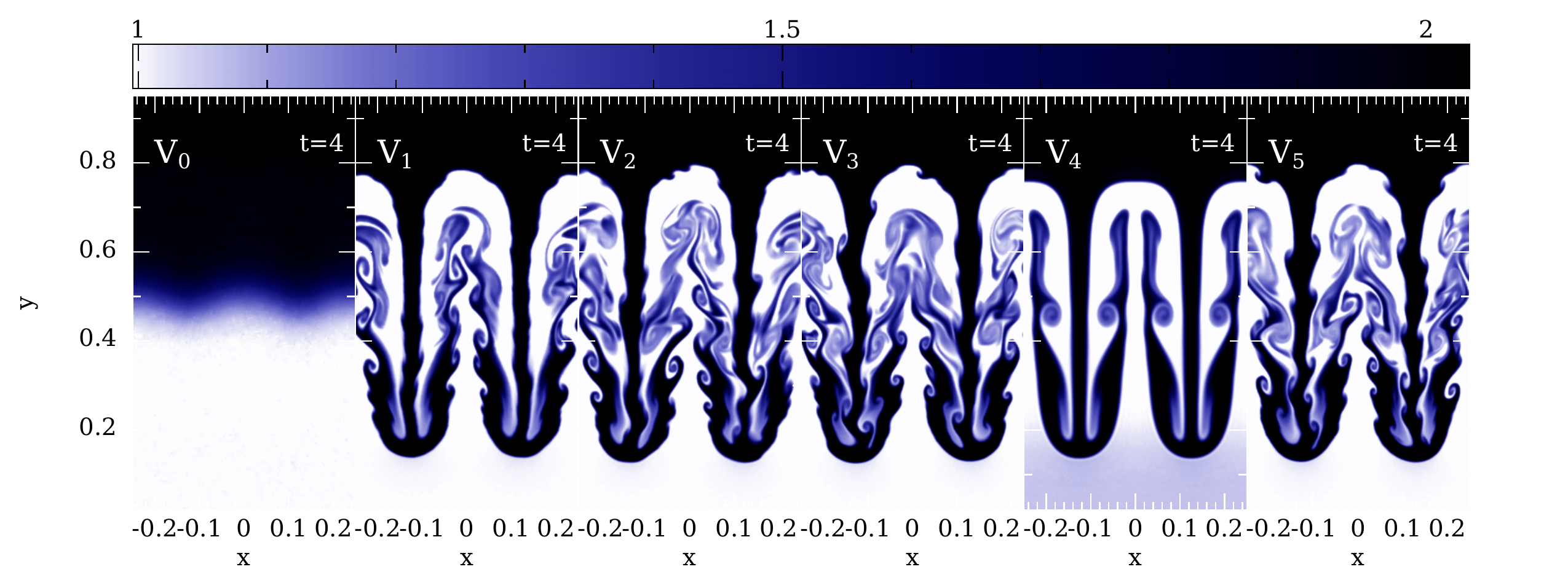} \\
   \caption{Density of Rayleigh-Taylor instability at $400^2$ for all versions.} 
   \label{fig:RaTa400}
\end{figure*}
\subsection{Rayleigh-Taylor test}
\label{sec:RaTa}
In a Rayleigh--Taylor instability test,  a density layer $\rho_t$ rests on top of a layer with
density $\rho_b < \rho_t$ in a constant acceleration field, e.g. due
to gravity.   When sinking down, the heavier fluid develops a
characteristic ``mushroom-like'' pattern.  Simulations with traditional
SPH implementations have  shown only retarded growth or even complete
suppression of the instability \citep{abel11,saitoh13}. 
We adopt again a quasi-2D setup, but evolve the fluid with a full 3D
code. For simplicity, we place 400 particles on a cubic lattice in the
$xy$-domain $[-0.25,0.25] \times [0,1]$ and use 20 layers of particles
in the $z$-direction, and also place 20 layers of particles as
boundaries around this core region. The properties of the particles below
$y=0$ are ``frozen'' at the values of the initial conditions, and we
multiply the temporal derivatives of particles with $y_a > 1$
with a damping factor 
\be
\xi= \exp\left\{-\left(\frac{y_a - 1}{0.05}\right)^2\right\},
\ee 
so that any evolution in this upper region is strongly suppressed. We
apply periodic boundaries in the $x-$direction 
at $x= \pm 0.25$. Similar to \citet{frontiere17} we use  $\rho_t=2$,
$\rho_b=1$, a constant acceleration $\vec{g}= -0.5 \hat{e}_y$ and 
\be
\rho(y)= \rho_b + \frac{\rho_t-\rho_b}{1+\exp[-(y-y_t)/\Delta]}
\ee
with transition width $\Delta=0.025$ and transition coordinate
$y_t=0.5$. We apply a small velocity perturbation to the interface 
\be
v_y(x,y)= \delta v_{y,0} [1 + \cos(8\pi x)][1 + \cos(5\pi(y-y_t))]
\ee
for $y$ in $[0.3,0.7]$ with an initial amplitude $\delta
v_{y,0}=0.025$, and we use a polytropic equation of state with
exponent $\Gamma=1.4$. The equilibrium pressure profile is given by
\be
P (y) = P_0 -  g \rho(y) [y - y_t]
\ee
with $P_0= \rho_t/\Gamma$, so that the sound speed is near unity in
the transition region.\\
The situation here is similar to the Kelvin-Helmholtz test: $V_0$ does 
hardly grow at all, the Riemann solver version $V_4$ grows slower
than the other good gradient versions and, again, maintains a very high 
degree of symmetry. The other versions develop filigrane fine structures
as they sink down. Once more, $V_3$ and $V_5$ show very similar 
performance, but are closely followed by the aLE-gradient version $V_2$.

%
\begin{figure*} 
   \centerline{
   \includegraphics[width=14cm]{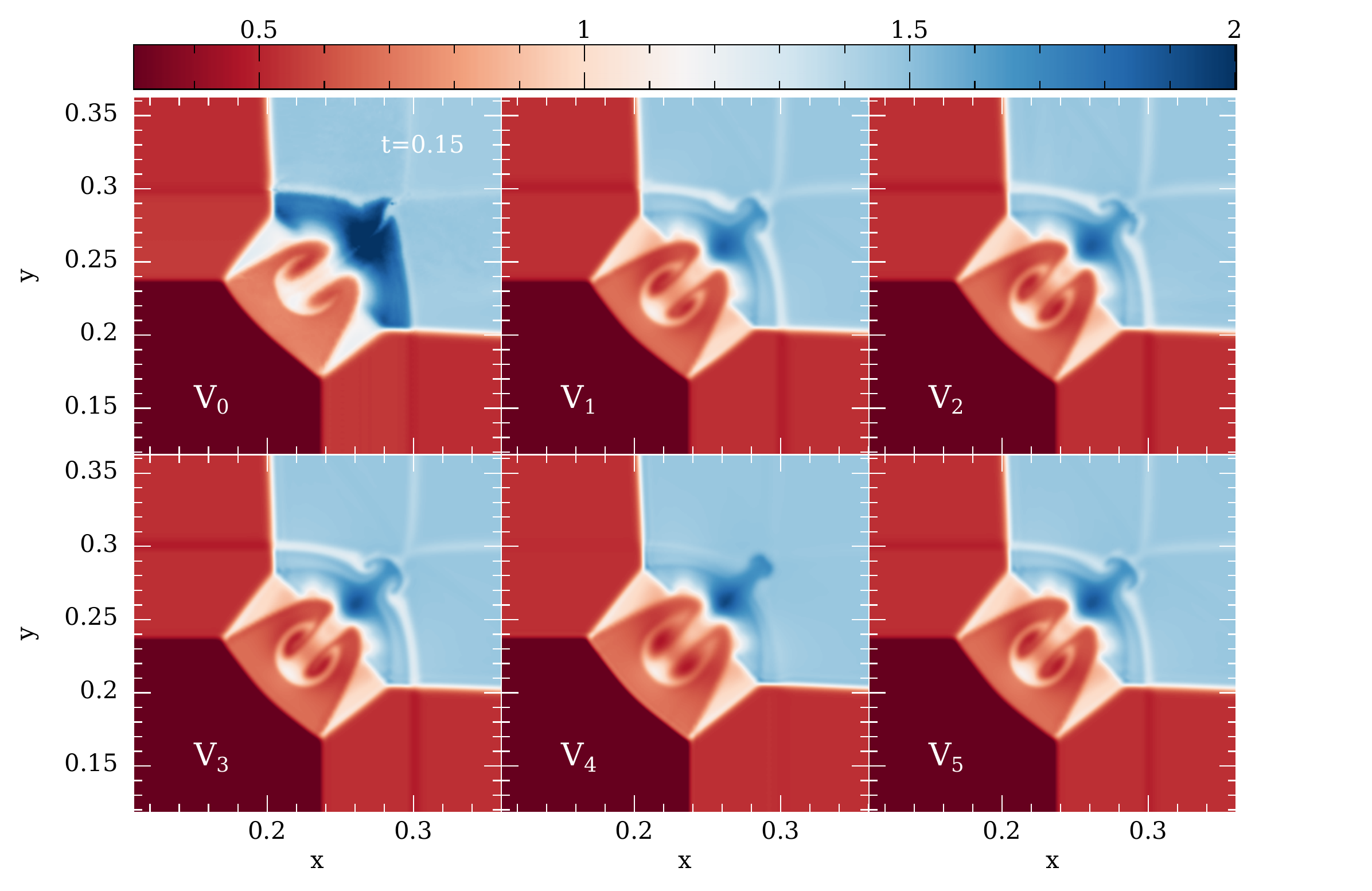} 
   }
   \caption{Schulz-Rinne test SR1 for all versions.}
   \label{fig:SR3}
\end{figure*}
%
\begin{figure*} 
   \centerline{
   \includegraphics[width=14.5cm]{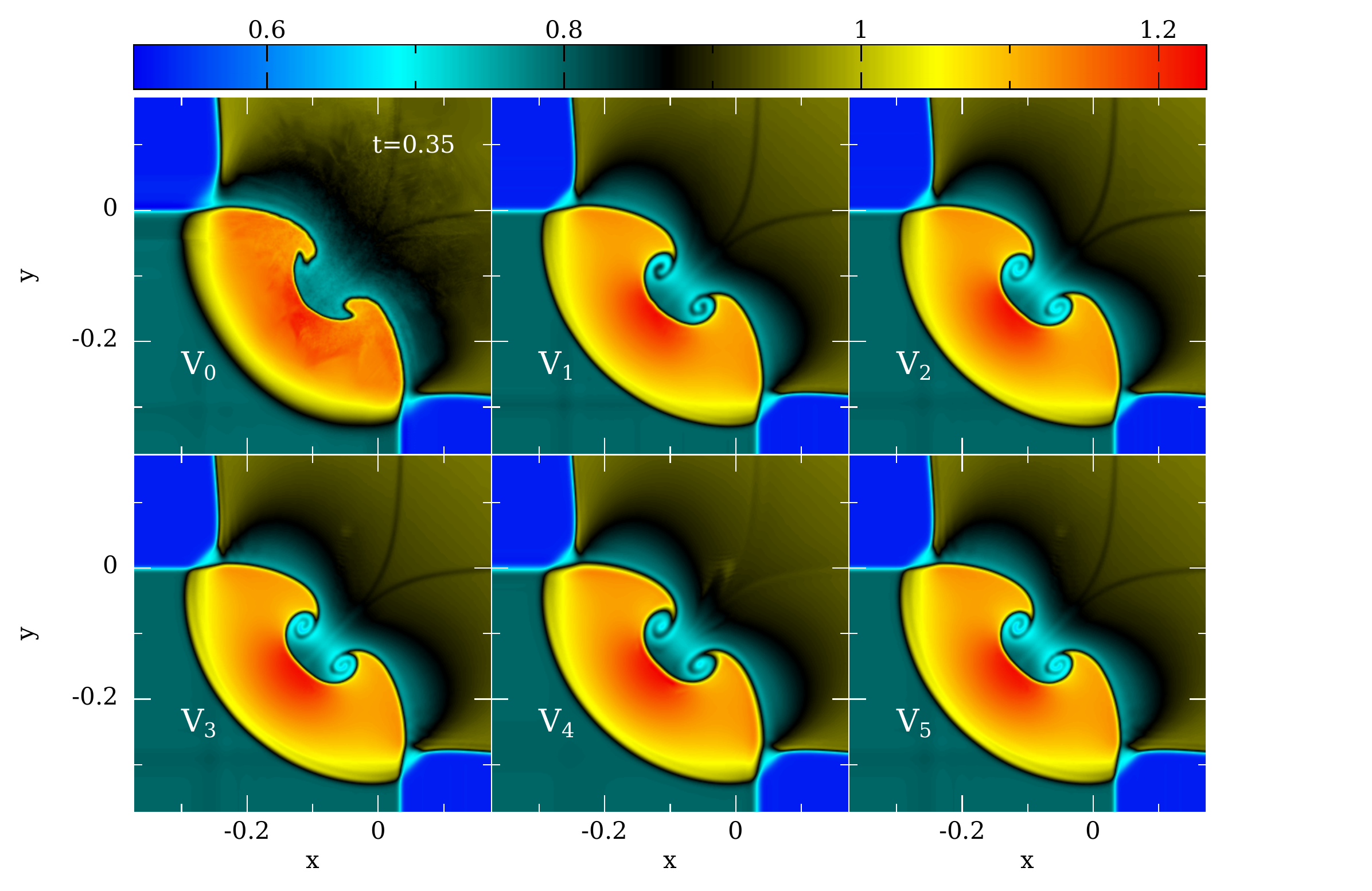} 
   }
   \caption{Schulz-Rinne test SR2 for all versions.}
   \label{fig:SR11}
\end{figure*}
%
\begin{figure*} %
   \centerline{
   \includegraphics[width=18cm]{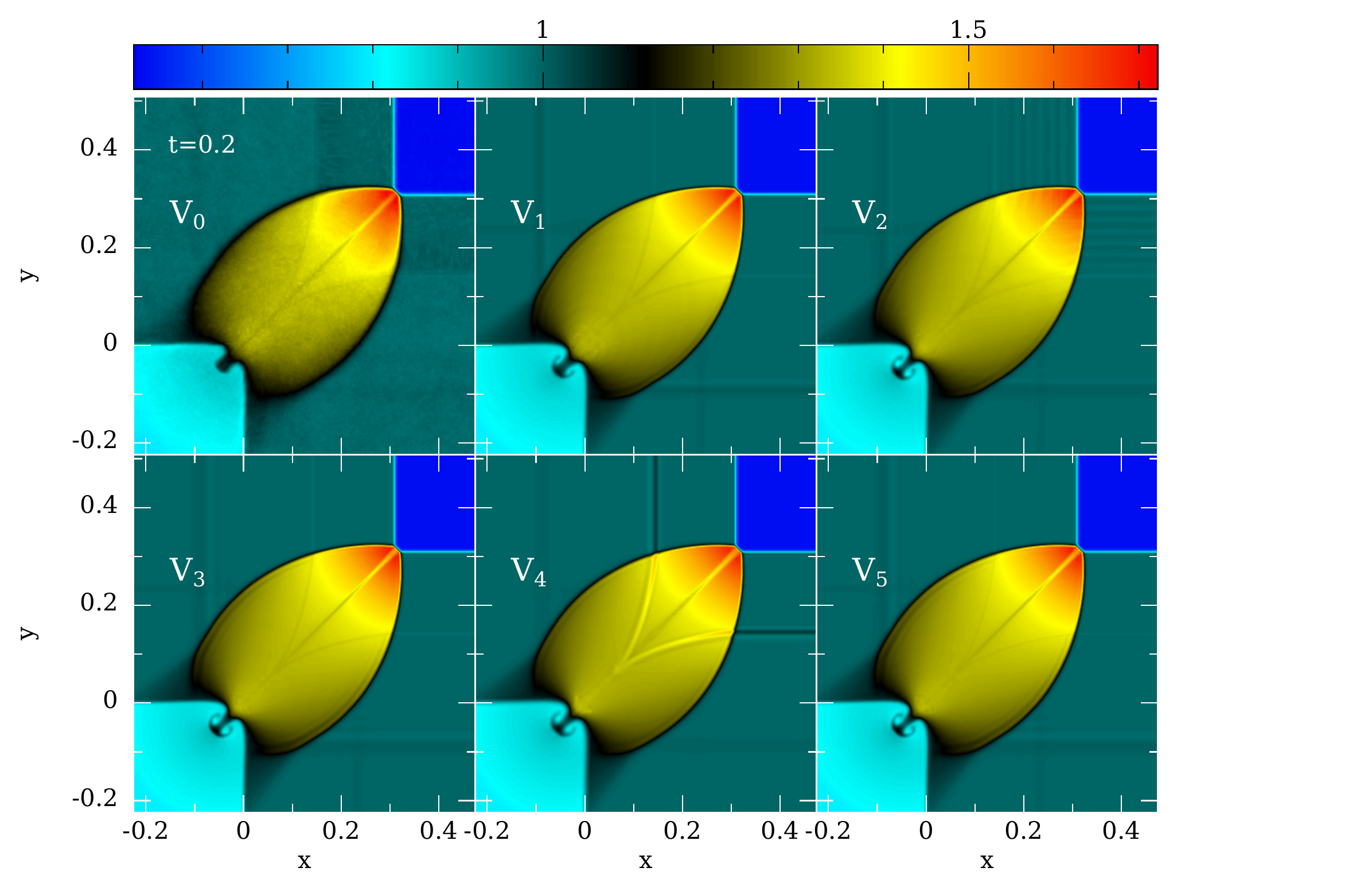} 
   }
   \caption{Schulz-Rinne test SR3 for all versions.}
   \label{fig:SR12}
\end{figure*}
\subsection{Schulz-Rinne tests}
\label{sec:SchulzRinne}
Schulz-Rinne tests \citep{schulzrinne93a} are very challenging 2D
benchmarks \citep{schulzrinne93a,lax98,kurganov02,liska03}.
They are constructed so that four constant states meet at one corner,
and the initial values are chosen so that an elementary wave, either a
shock, a rarefaction, or a contact discontinuity, appears at each
interface.  During subsequent evolution, complex wave patterns
emerge for which exact solutions are not known. Such tests have not
often been shown for SPH. We are only aware of the work of
\citet{puri14}, applying Godunov SPH with approximate Riemann solvers,
the tests in the \Ma code \citep{rosswog20a} and in a Riemann solver
approach with reproducing kernels \citep{rosswog25a}.\\
\noindent Here, we show three such tests. We place particles on a cubic
lattice in a 3D slice thick enough so that the mid-plane is unaffected
by edge effects (we use 20 particle layers in $z$-direction),   
so that 300 x 300 particles are within  $[x_c-0.3,x_c+0.3] \times [y_c-0.3,y_c+0.3]$, 
where $(x_c,y_c)$ is the contact point of the quadrants, and we use
a polytropic exponent $\Gamma=1.4$ in all tests. We refer to these
Schulz-Rinne type problems as SR1-SR3 \footnote{In 
  \citet{kurganov02} these are tests have the numbers 3, 11, and 12.} and give their initial
parameters for each quadrant in Table~\ref{tab:SchulzRinne}. \\
\begin{table}
\caption{Initial conditions for the Schulz-Rinne-type 2D Riemann problems.}
\begin{tabular}{ l c | c | c | c | c | c | c |}
\hline
 & & SR1; contact point: $(0.3,0.3)$ & &\\
  \hline	
  \hline		
  variable & NW & NE & SW & SE \\ \hline
  $\rho$ &  0.5323 &  1.5000  & 0.1380 &  0.5323 \\
  $v_x$ &  1.2060 &  0.0000   & 1.2060 & 0.0000  \\
  $v_y$ &   0.0000 & 0.0000   & 1.2060 & 1.2060 \\
  $P$    &   0.3000 & 1.5000   & 0.0290 & 0.3000    \\
  \hline  
   & & SR2; contact point: $(0.0,0.0)$ & &\\
  \hline	
  \hline		
  variable & NW & NE & SW & SE \\ \hline
  $\rho$ &  0.5313 &  1.0000 & 0.8000  &  0.5313 \\
  $v_x$ &   0.8276 &  0.1000 & 0.1000 & 0.1000  \\
  $v_y$ &   0.0000 & 0.0000  & 0.0000  &  0.7276\\
  $P$    &   0.4000  & 1.0000  & 0.4000 & 0.4000    \\
  \hline  
   & & SR3; contact point: $(0.0,0.0)$ & &\\
  \hline	
  variable & NW & NE & SW & SE \\ \hline
  $\rho$ &  1.0000 &  0.5313 & 0.8000  &  1.000 \\
  $v_x$ &   0.7276 &  0.0000 & 0.0000 & 0.0000  \\
  $v_y$ &   0.0000 & 0.0000  & 0.0000  &  0.7262\\
  $P$    &   1.0000  & 0.4000  & 1.0000 & 1.0000    \\
  \hline  
\end{tabular}
\label{tab:SchulzRinne}
\end{table}
\noindent For the first test, see Fig.~\ref{fig:SR3}, $V_3$ and $V_5$ perform best, but closely followed
by $V_2$. They show a high degree of symmetry and "mushroom-like" structures in both forward 
an backward direction. The Riemann solver version $V_4$ seems to suppress at least the "backward 
mushroom" (at $x\approx 0.3$ and $y\approx 0.3$) and the "forward mushroom"  (at $x\approx 0.25$ 
and $y\approx 0.25$) is not as crisp as for $V_3$ and $V_5$. The gradient accuracy is important
here for maintaining the symmetry, both standard kernel gradient versions ($V_0$ and $V_1$) show
noticeable asymmetries.\\
The situation is very similar for the second test, see Fig.~\ref{fig:SR11}. Again $V_0$ hardly shows much structure,
$V_1$ is already substantially better, $V_4$ is reluctant to curl in on the "rim of the mushroom", $V_3$ and $V_5$
are best, but closely followed by $V_2$.\\
Practically the same can be said about the third Schulz-Rinne test, see Fig.~\ref{fig:SR12}. Once more,
$V_3$ and $V_5$ perform best, but closely followed by $V_2$, $V_4$ shows a delayed formation of the
mushroom-like structure which is essentially suppressed in $V_0$. It is also worth pointing out  the
spurious wave pattern that emerges in the case of $V_2$, see e.g. the region $x>0.3$ and $y> 0.1$.

\section{Summary}
\label{sec:summary}
In this study we explored the impact of the gradient accuracy on the performance of various
SPH formulations that are specified by the choice of a scalar $\Psi$, see Eqs.~(\ref{eq:drhodt_psi}) to 
(\ref{eq:dudt_psi}). Consistent with earlier studies, we found that $\Psi= \rho$ is an excellent 
choice and therefore we used this as a baseline model. Apart from  the standard 
SPH kernel gradient we explored  an approximation to the "linearly exact gradient" (aLE-gradients) and linearly 
reproducing kernel (RPK) gradients. The aLE-gradients become linearly exact in the limit of an exact partition 
of unity and they only cost the inversion of a small $3\times3$ matrix. The RPK kernels
are constructed so that they exactly fulfill the discrete linear consistency relations, see Eq.~(\ref{eq:consistency_relations}),
and their construction comes at a  noticeable computational cost, for their explicit expressions see Appendix \ref{sec:RPK}.  
Both aLE and RPK approaches massively improve the gradient accuracy, but while ensuring exact energy and 
momentum conservation, they sacrifice the exact conservation of angular momentum since the inter-particle forces 
can no longer be guaranteed to be exactly along the line connecting two particles.\\
In all but one SPH formulations we used shock dissipation, i.e. artificial viscosity and conductivity, 
and in the "jump terms" which are responsible for the dissipation, we use linearly reconstructed 
quantities limited by the van Albada slope limiter \citep{vanAlbada82}. In addition to the slope-limited reconstruction, we steer the dissipation
parameters, so that they decay to zero where they are not needed. The dissipation parameters 
are raised as indicated by a Cullen-Dehnen-type (CD) shock indicator reacting on $\d(\nabla \cdot \vec{v})/dt$, 
but we also add a sensitive trigger that measures local noise in the particle distribution.
We compared these modern SPH formulations with a recent particle  hydrodynamics 
approach \citep{rosswog25a} that also uses RPKs, but  applies Roe's Riemann solver 
instead of shock dissipation.\\
Our main results can be summarized as follows:
\bi
\i Not too surprisingly, a major improvement compared to "old school" SPH ($V_0$) with constant 
   dissipation and cubic spline kernels comes from the reduction of unwanted dissipation. The
   $V_0$ formulation gives reasonable results in shocks, but massively underperforms in
   all tests related to instabilities.
\i Shock tests are not very sensitive to the gradient accuracy, only in Riemann problem 2, see Fig.~\ref{fig:Riemann2}, 
we find somewhat noisy inner regions for all but the RPK gradients. But we find that the gradient accuracy
plays a major role for all tests that involve fluid instabilities, see Sec.~\ref{sec:KeHe}, \ref{sec:RaTa} 
and \ref{sec:SchulzRinne}.
\i As expected, the numerically most expensive gradients using reproducing kernels perform
   best, but they are closely followed by the substantially cheaper aLE gradients. The differences
   between the two show up mostly in the instability and the Schulz-Rinne tests. We also find the RPK
   gradients consistently trigger lower dissipation values, see Fig.~\ref{fig:triggered_dissipation}, 
   indicating a more accurate particle motion on the length scales of the kernel support.
\i While the CD-type trigger works very well in shocks, we find that it does not trigger sufficient dissipation in
   the complex Schulz-Rinne tests, see the left column in Fig.~\ref{fig:effect_noise_trigger}. But together with
    our noise trigger it provides an appropriate amount of dissipation, see 
   the right column in  Fig.~\ref{fig:effect_noise_trigger}, and without triggering too much dissipation in the rather
   gentle fluid instability tests. We find this combination of shock and noise triggers accurate and robust. 
\i Comparing the shock dissipation approaches with the Riemann solver case, we find interesting
   differences. While the Riemann solver approach ($V_4$ in Sec.~\ref{sec:SPH_formulations})
   has some advantage in the shock tests, producing the smallest amount of oscillations (although closely 
   followed by $V_3$ and $V_5$), it produces more dissipation
   in  rarefaction regions than the shock dissipation approach. This is simply because the Riemann 
   solver dissipation is applied independent of whether the local flow is converging or diverging 
   while the shock dissipation is only applied for converging flows. More interestingly, the Riemann 
   solver approach shows some resistance for 
   fluid instabilities to develop. In the fluid instability tests it enforces nearly perfect symmetry, but 
   the instabilities grow substantially slower than in the shock dissipation cases. In fact, at very low 
   resolution ($64^2$), only the "old school" approach $V_0$ behaves worse than the Riemann solver
   approach $V_4$, see Fig.~\ref{fig:mode_growth}.
   \ei
The RPK approach is clearly a major improvement over standard SPH gradients. The aLE-gradients
are less accurate, but we still find them very useful, mostly because of their simple implementation and negligible
computational cost and in many of the shown tests they are nearly as good as the RPK approach.
 Based on this test suite, we consider our SPH version $V_3$ as the best one,
but very closely followed by $V_5$. The simpler $V_2$ version is still a substantial improvement over
the standard SPH gradients. The standard kernel gradient SPH approach with good choices, $V_1$, 
performs overall still pretty well. Due its reluctance to let instabilities grow, we rank the Riemann solver
version $V_4$ lower than the previous versions, despite its very good performance in shocks.  While the best of 
these versions represent the current state-of-the-art for SPH-type hydrodynamics, further improvements 
of particle hydrodynamics is probably possible and should be explored in future work.

\section*{Acknowledgements}
SR has been supported by the Swedish Research Council (VR) under 
grant number 2020-05044, by the research environment grant
``Gravitational Radiation and Electromagnetic Astrophysical
Transients'' (GREAT) funded by the Swedish Research Council (VR) 
under Dnr 2016-06012, by the Knut and Alice Wallenberg Foundation
under grant Dnr. KAW 2019.0112,  by the Deutsche 
Forschungsgemeinschaft (DFG, German Research Foundation) under 
Germany's Excellence Strategy - EXC 2121 ``Quantum Universe'' - 390833306 
and by the European Research Council (ERC) Advanced 
Grant INSPIRATION under the European Union's Horizon 2020 research 
and innovation programme (Grant agreement No. 101053985).\\
The simulations for this paper have been performed on the facilities of
 North-German Supercomputing Alliance (HLRN), and at the SUNRISE 
 HPC facility supported by the Technical Division at the Department of 
 Physics, Stockholm University, and on the HUMMEL2 cluster funded 
 by the Deutsche Forschungsgemeinschaft  (498394658). Special thanks 
 go to Mikica Kocic (SU),  Thomas Orgis and Hinnerk Stueben (both UHH) for their 
 excellent support.\\
 Some  plots have been produced with the software \texttt{splash} \citep{price07d}.
  

\appendix

\section{Explicit expressions for reproducing kernels}
\label{sec:App_RPK}
The gradient of the kernel $\mathcal{W}$ with respect to position $\vr_a$,
$\p_k \mathcal{W}_{ab}\equiv (\nabla_a)^k  \mathcal{W}_{ab}$,
 reads
\begin{align}
\p_k \mathcal{W}_{ab}=& \; A_a \;
B_a^k \; \bar{W}_{ab} + A_a \left(1+ B_a^i (\vec{r}_{ab})^i\right)                                                                                                    
\nabla_a^k \bar{W}_{ab} + \nonumber \\
&
\left(  1+  B_a^i  (\vec{r}_{ab})^i \right)
\bar{W}_{ab} (\nabla_a^k A)_a + A_a \;
(\vec{r}_{ab})^i \; (\nabla_a^k B)_a^i \;
\bar{W}_{ab}.
\label{eq:RPK_kernel_gradient}
\end{align}
Taking the gradient of $\mathcal{W}_{ba}$ with respect to $\vr_b$,
$\p_k \mathcal{W}_{ba}\equiv (\nabla_b)^k  \mathcal{W}_{ba}$,
results in
\begin{align}
\p_k \mathcal{W}_{ba}=& \; A_b
\; B_b^k \; \bar{W}_{ab} - A_b \left(1 - B_b^i (\vec{r}_{ab})^i\right) \nabla_a^k \bar{W}_{ab} + \nonumber \\
& \left( 1- B_b^i (\vec{r}_{ab})^i \right) \bar{W}_{ab} (\nabla_b^k A)_b - A_b \; (\vec{r}_{ab})^i \; (\nabla_b^k B)_b^i  \; \bar{W}_{ab} ,    
\end{align}
where we have used $\bar{W}_{ab}= \bar{W}_{ba}$, $\vec{r}_{ba}=
-\vec{r}_{ab}$ and $\nabla_b^k \bar{W}_{ba}= - \nabla_a^k
\bar{W}_{ab}$.
The derivations that lead to the explicit expressions for
$A_a$, $B_a$ and their derivatives are straight-forward, but lengthy.
One needs the discrete moments at position $a$
\bea
(M_0)_a &\equiv& \sum_b V_b \; \bar{W}_{ab} \\
(M_1^i)_a &\equiv& \sum_b (\vec{r}_{ab})^i \; V_b \;  \bar{W}_{ab} \\
(M_2^{ij})_a &\equiv& \sum_b (\vec{r}_{ab})^i  \; (\vec{r}_{ab})^j  \; V_b \; \bar{W}_{ab}
\eea
and their derivatives
\be
\hspace*{-5.5cm}(\p_k M_0)_a = \sum_b V_b \; \nabla_a^k  \bar{W}_{ab} 
\ee
\be
\hspace*{-3cm}(\p_k M_1^i)_a = \sum_b V_b \left[ (\vec{r}_{ab})^i (\nabla_a^k \bar{W}_{ab}) + \delta^{ki} \bar{W}_{ab} \right]
\ee
\be
(\p_k M_2^{ij})_a  =  \sum_b V_b \; \left[ (\vec{r}_{ab})^i  \; (\vec{r}_{ab})^j  \; (\nabla_a^k \bar{W}_{ab}) + 
(\vec{r}_{ab})^i \; \delta^{jk} \; \bar{W}_{ab} +  (\vec{r}_{ab})^j \; \delta^{ik} \; \bar{W}_{ab} \right].
\ee
With these expressions at hand,  one can straight forwardly calculate the kernel parameters
\bea
A_a &=& \frac{1}{(M_0)_a - (M_2^{ij})_a^{-1} \; (M_1^i)_a \; (M_1^j)_a}\\
B_a^i&=& - (M_2^{ij})_a^{-1} \; (M_1^j)_a
\eea
and their somewhat lengthy, but otherwise straight forwardly calculable, derivatives which are 
needed for $\p_k \mathcal{W}_{ab}$, see Eq.~(\ref{eq:RPK_kernel_gradient})
\begin{align}
\p_k A_a =& -A^2_a \left[ (\p_k M_0)_a - 2 (M_2^{ij})_a^{-1}   (M_1^j)_a  \left( \p_k M_1^i \right)_a   \right. \nonumber\\
 & \left. + (M_2^{il})_a^{-1}   (\p_k M_2^{lm})_a (M_2^{mj})_a^{-1} (M_1^j)_a (M_1)_a^i \right]
\end{align}
and
\be
\p_k B_a^i= -(M_2^{ij})_a^{-1}  \; (\p_k M_1^j)_a + (M_2^{il})^{-1} \; (\p_k M_2^{lm})_a  (M_2^{mj})_a^{-1} (M_1^j)_a. 
\ee

\section{Comparison for different common choices of $\Psi$}
\label{sec:choice_of_Psi}
We had given in Sec.~\ref{sec:SPH_formulations} SPH formulations that contain a scalar field $\Psi$
and we want to show here a concise comparison between three common choices. Each time we use
the same density summation, Eq.~(\ref{eq:dens_sum_Wab}), and we use the gradients $(\nabla \mathcal{W})_{ab}$,
see Eq.~(\ref{eq:RPK_gradient}). For $\Psi=1$ one recovers (apart from using the RPK gradients) the "vanilla
ice" symmetrization of SPH that was also used in version $V_0$. For $\Psi= \rho/\sqrt{P}$ one recovers the 
\cite{hernquist89} symmetrization
\bea
\frac{d\vec{v}_a}{dt}&=& - \sum_b m_b  \left( 2\frac{\sqrt{P_a  P_b}}{\rho_a
    \rho_b} \right) (\nabla \mathcal{W})_{ab}\\
\frac{d u_a}{dt}        &=& \sum_b m_b \left(\frac{\sqrt{P_a P_b}}{\rho_a \rho_b}\right)
\vec{v}_{ab} \cdot (\nabla \mathcal{W})_{ab},
\eea
and for $\Psi=\rho$ one finds our default symmetrization that is used in the SPH formulations
$V_1$, $V_2$ and $V_3$. \\
Here we use each time  exactly the same shock dissipation, see Sec.~\ref{sec:shock_diss},
kernel, initial data and gradients, the only change is the {\em symmetry} in the equations.
We illustrate the differences  with one instability, the Kelvin-Helmholtz instability, and a shock
with strong density gradients, a Sedov explosion test. For both tests we have reference solutions to compare with
and both tests are  set up exactly as described in the main text.\\
We perform the Kelvin-Helmholtz test at a resolution of only $64^2$ so that potential differences should be clearly visible.
\begin{figure} 
   \centerline{
   \includegraphics[width=9cm]{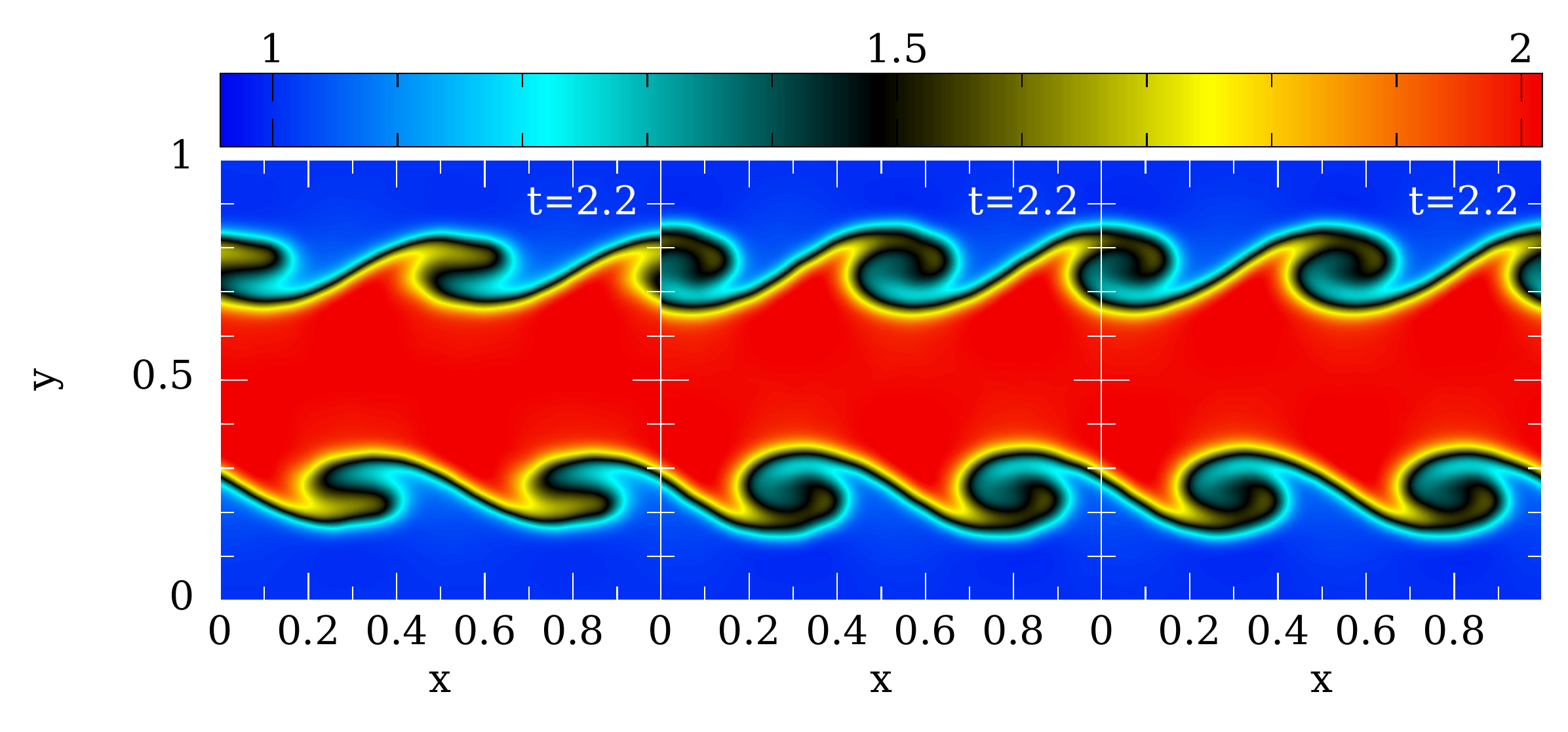} 
   }
   \caption{Comparison of a low resolution ($64^2$) Kelvin-Helmholtz test for different choices
   of the free scalar $\Psi$. The left panel shows the results for "vanilla ice SPH" ($\Psi=1$), the middle for
   the Hernquist and Katz symmetrization ($\Psi=\rho/\sqrt{P}$), and the right panel uses our default choice  
   ($\Psi=\rho$). Clearly, the "vanilla ice" version grows slower than the other two.}
   \label{fig:KeHe_Psi_comp}
\end{figure}
%
\begin{figure} 
   \centerline{
   \includegraphics[width=9cm]{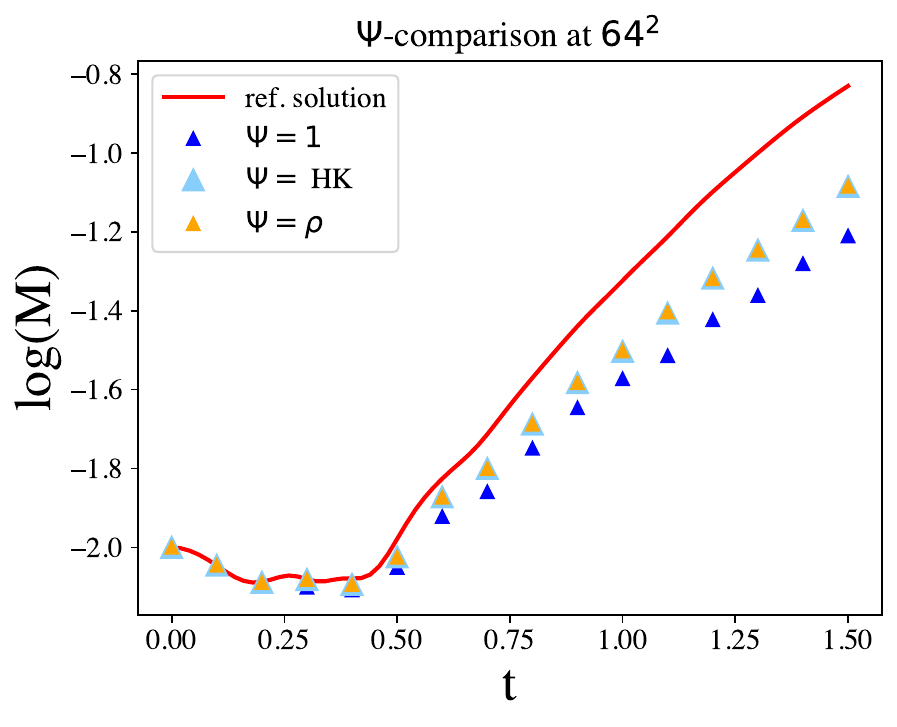} 
   }
   \caption{Comparison of a low resolution ($64^2$) Kelvin-Helmholtz test for the different choices
   of the free scalar $\Psi$ that are shown in Fig.~\ref{fig:KeHe_Psi_comp}. Note that the symbol for 
   the Hernquist and Katz symmetrization ("HK") has  been increased since otherwise it would not be 
   visible below the $\Psi=\rho$ results.}
   \label{fig:KeHe_Psi_comp_growth}
\end{figure}
For the Kelvin-Helmholtz test the vanilla ice symmetrization ($\Psi=1$) clearly grows slower than the other two
which perform here similarly, see Fig.~\ref{fig:KeHe_Psi_comp}.  On close inspection one finds that the $\Psi=\rho$ 
symmetrization grows fastest, which is also visible in the measured growth rates, see Fig.~\ref{fig:KeHe_Psi_comp_growth}.\\
In Fig.~\ref{fig:Sedov_Psi_comp} we zoom in on the density peak of a $128^2$ Sedov test for all three versions. Also here, the
$\Psi= \rho$ performs best, now having a clear advantage  over the Hernquist-Katz symmetrization.
\begin{figure} 
   \centerline{
   \includegraphics[width=9cm]{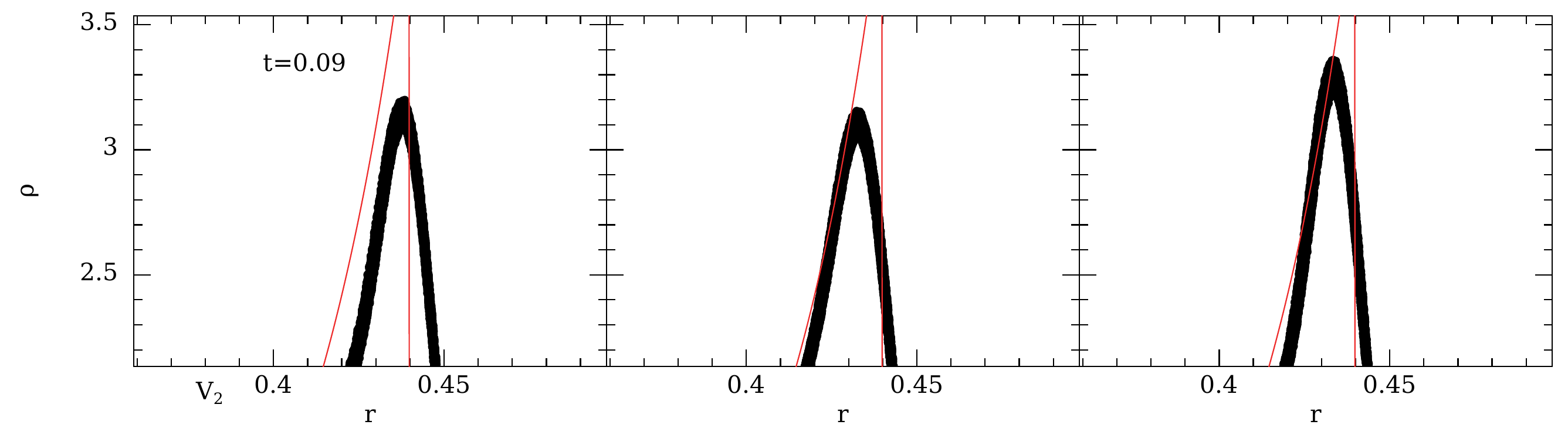} 
   }
   \caption{Comparison of the peak heights reached in the Sedov explosion test ($128^2$; all particles plotted) 
   for different choices of the free scalar: $\Psi=1$ (left panel, "vanilla ice"), $\Psi= \rho/\sqrt{P}$ (middle, Hernquist-Katz) 
   and $\Psi= \rho$ (right panel; our default choice). Here  $\Psi=\rho$ clearly reaches the largest peak height.}
   \label{fig:Sedov_Psi_comp}
\end{figure}
In summary, from the explored choices $\Psi= \rho$ has a clear advantage over the other possibilities. We are not aware of
any disadvantage of this formulation and therefore use it as our default.

\section{Effectiveness of the noise trigger for dissipation}
\label{sec:effectiveness_noise_trigger}
\begin{figure*} 
   \centerline{\includegraphics[width=14cm]{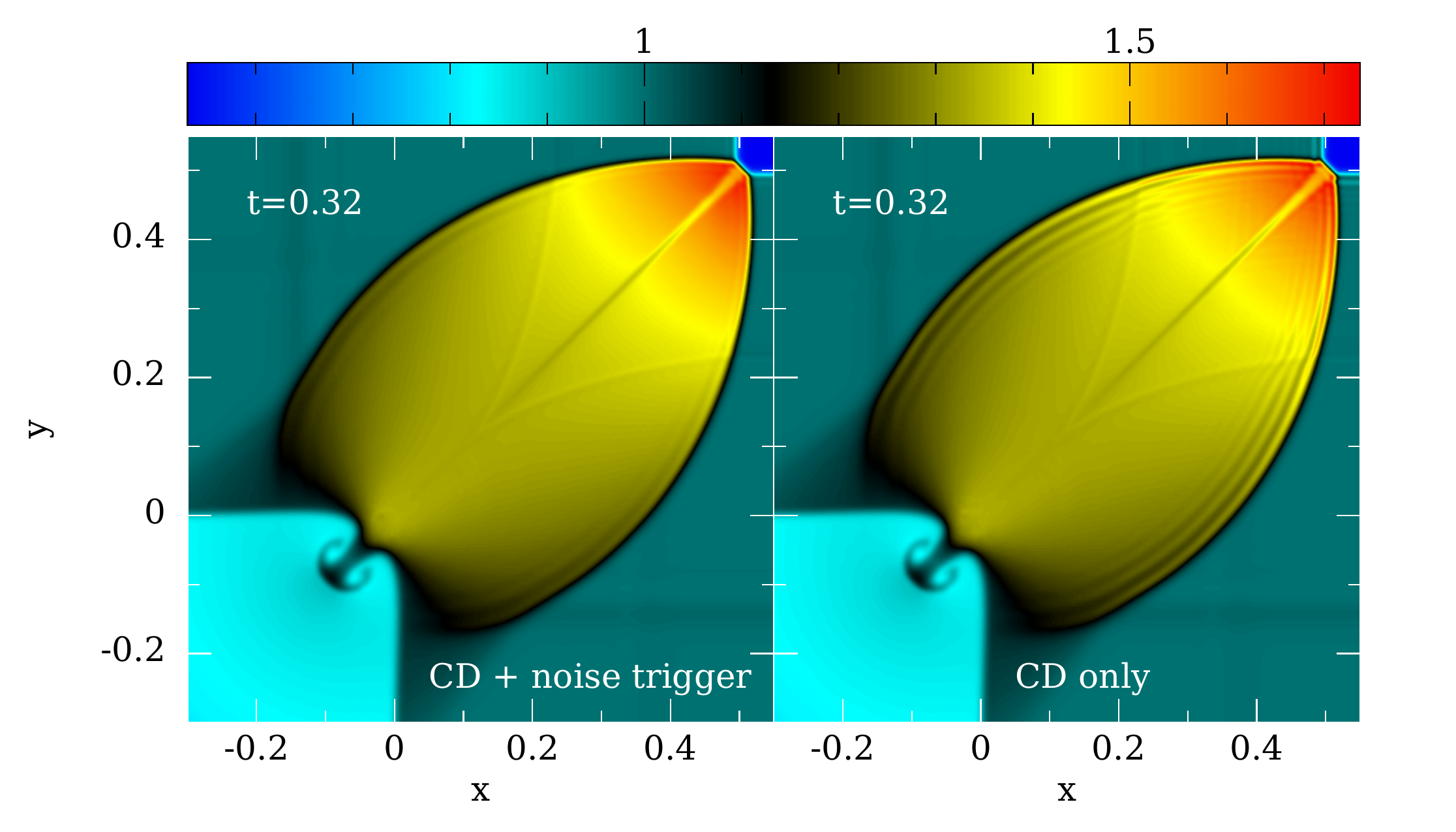}}
   \vspace*{-0.2cm}
   \centerline{\includegraphics[width=14cm]{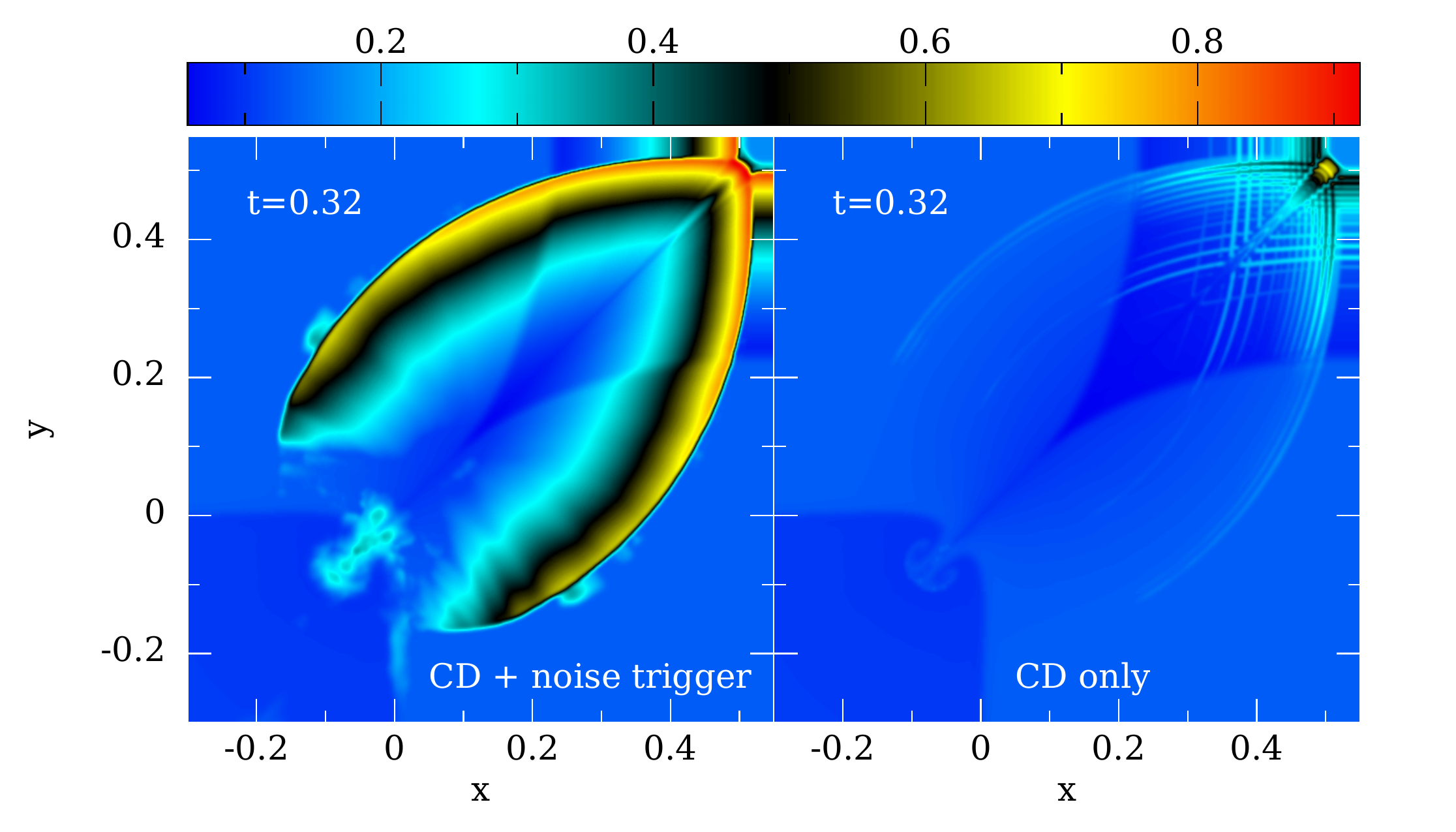}}
   \caption{Comparison of a lower resolution ($200^2$) simulation of Schulz Rinne test 3, 
   once using the Cullen-Dehnen-type (CD) trigger as described in 
   the main text together with our noise trigger (left column) and once relying exclusively on the CD trigger.
   The upper row shows density, the lower one the dissipation parameter $\alpha$.
   Clearly the use of noise trigger efficiently suppresses the spurious waves that are visible the "CD only" case.}
   \label{fig:effect_noise_trigger}
\end{figure*}

We want to briefly illustrate the impact of the noise trigger, see Eq.~(\ref{eq:noise_trigger}). The functional
form of this trigger was suggested by \cite{rosswog15b}, though in a special-relativistic context.
Contrary to the original suggestion, we choose here a rather small prefactor (=0.01) in the reference value, 
see Eq.~(\ref{eq:alpha_noise}). This reference value decides on how to translate the trigger $T^n_a$ 
into a dissipation parameter value, $\alpha_{a, \rm noise}$.\\
To illustrate that such an additional triggering on noise is really desirable and that the suggested
trigger is really effective, we run a lower-resolution simulation ($200^2$) of the Schulz-Rinne test 3
with SPH version $V_3$,  once  using only the \cite{cullen10} (CD) trigger  as described in Sec.~\ref{sec:shock_diss} and once using 
in addition our noise trigger, see Fig.~\ref{fig:effect_noise_trigger}. In the CD-only case not sufficient
dissipation is triggered and a number of spurious wiggles appear (right column) which are efficiently suppressed
by the suggested noise trigger (left column).\\ 
Note that while the Wendland C4 kernel is the best overall compromise, in this test a high order
$WH_8$ kernel \cite{cabezon08} does not even show the small post-shock wiggle that is still visible
in the upper left panel.
%
\begin{figure*} 
   \centerline{\includegraphics[width=14cm]{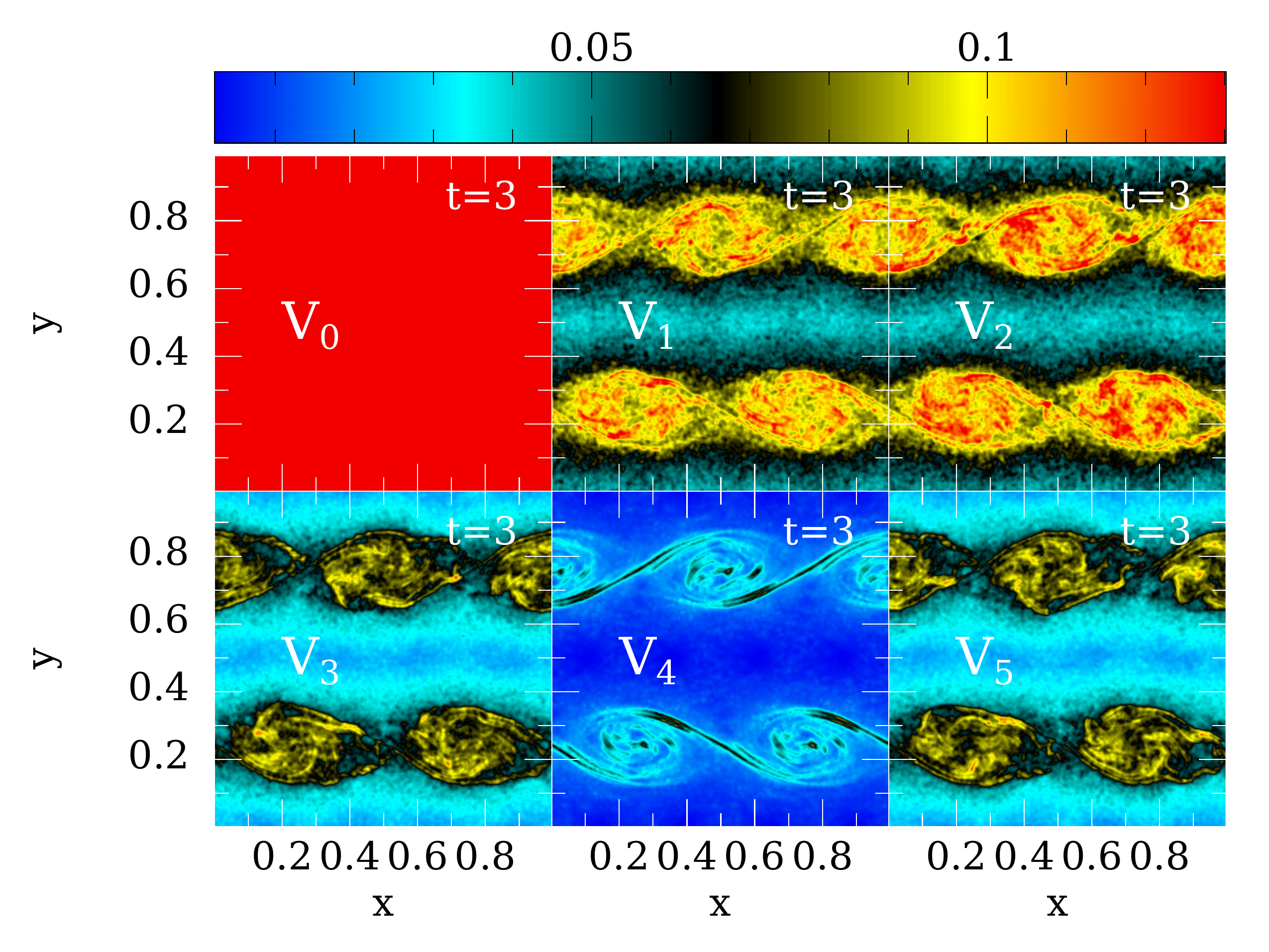}}
   \caption{Values of the dissipation parameter $\alpha$ for the different versions in a Kelvin-Helmholtz test ($512^2$). 
   For visibility reasons  we cut the colour bar at a maximum of 0.13. Note that for $V_4$ the dissipation parameter is triggered
   exactly as in the other cases, but it is not used, since  the dissipation comes from the Riemann solver.
   The $\alpha$-value however can still indicate the noise on the kernel level. Note that the RPK gradients ($V_3,V_4,V_5$)
   trigger noticeably less dissipation than either kernel gradient ($V_1$) or the aLE gradients ($V_2$).}
   \label{fig:triggered_dissipation}
   \end{figure*}

\bibliographystyle{mn2e}
\bibliography{astro_SKR.bib}
\end{document}